# Origin of interface limitation in CuInS$_2$ – based solar cells

Mohit Sood*[1], Jakob Bombsch*[2], Alberto Lomuscio[1], Sudhanshu Shukla[1], Alberto Lomuscio[1], Claudia Hartmann[2], Johannes Frisch[2,3], Wolfgang Bremsteller[2,3], Shigenori Ueda[4,5,6], Regan G. Wilks[2,3], Marcus Bär[2,3,7,8], Susanne Siebentritt[1]

*equal contributions

(E-mail addresses: mohit.sood@uni.lu, jakob.bombsch@helmholtz-berlin.de)

[1]*Laboratory for Photovoltaics, Department of Physics and Materials Science, University of Luxembourg, Belvaux, L-4422, Luxembourg*

[2]*Department Interface Design, Helmholtz-Zentrum Berlin für Materialien und Energie GmbH (HZB), 12489 Berlin, Germany*

[3]*Energy Materials In-situ Laboratory Berlin (EMIL), Helmholtz-Zentrum Berlin für Materialien und Energie GmbH, 12489 Berlin, Germany*

[4]*NIMS Synchrotron X-ray Station at SPring-8, National Institute for Materials Science (NIMS), 1-1-1 Kouto, Sayo, Hyogo, 679-5148 Japan*

[5]*Research Center for Advanced Measurement and Characterization, NIMS, 1-2-1 Sengen, Tsukuba, Ibaraki 305-0047, Japan*

[6]*Research Center for Functional Materials, NIMS, 1-1 Namiki, Tsukuba, Ibaraki 305-0044, Japan*

[7]*Department X-ray Spectroscopy at Interfaces of Thin Films, Helmholtz Institute for Renewable Energy (HI ERN), 12489 Berlin, Germany*

[8]*Department of Chemistry and Pharmacy, Friedrich-Alexander Universität Erlangen-Nürnberg (FAU), 91058 Erlangen, Germany*

**Abstract**

Copper indium disulfide ($CuInS_2$) grown under Cu-rich conditions exhibits high optical quality but suffers predominantly from charge carrier interface recombination resulting in poor solar cell performance. An unfavorable "cliff"-like conduction band alignment at the buffer/$CuInS_2$ interface could be a possible cause of enhanced interface recombination in the device. In this work, we exploit direct and inverse photoelectron spectroscopy together with electrical characterization to investigate the cause of interface recombination in $Zn(O,S)/CuInS_2$ devices. Temperature-dependent current-voltage analysis indeed reveal an activation energy of the dominant charge carrier recombination path, considerably smaller than the absorber bandgap, confirming the dominant recombination channel to be present at the $Zn(O,S)/CuInS_2$ interface. However, photoelectron spectroscopy measurements indicate a small "spike"-like conduction band offset of 0.1 eV at the $Zn(O,S)/CuInS_2$ interface, excluding an unfavorable energy level alignment to be the prominent cause for strong interface recombination. The observed band bending upon interface formation also rules out Fermi level pinning as the main reason, leaving near-interface defects (as recently observed in Cu-rich $CuInSe_2$)[1] as the likely reason for the performance limiting interface recombination.

## INTRODUCTION

Copper indium disulfide (CuInS$_2$), particularly grown under Cu-rich conditions, has been extensively studied in the past and shows superior opto-electronic properties, compared to its Cu-poor counterpart.[2-5] However, the recombination of charge carriers at the buffer/absorber interface remains a bottleneck for devices realized with CuInS$_2$ grown under Cu-rich conditions.[6] This prevents the transfer of the good opto-electronic absorber properties into a high-efficiency solar cell. Specifically, interface recombination remains a major hurdle in achieving high open-circuit voltages (V$_{OC}$).[7] In devices prepared with a conventional CdS buffer layer, a negative conduction band offset (CBO), *i.e.* a "cliff" in the conduction band at the buffer/absorber (CdS/CuInS$_2$) interface is suggested to be the prominent reason for interface recombination.[8-10] Therefore, alternative buffer layers such as Zn(O,S)[11], Zn(Se,O),[12] and In(OH,S)[13] have been employed in the past. Among these, Zn(O,S) buffer layers remain the most promising candidates, already resulting in devices with a maximum power conversion efficiency (PCE) of 11%.[11] In the previous energy level alignment study of the Zn(O,S)/CuInS$_2$ interface, we found the presence of a significant positive (*i.e.*, "spike"-like) CBO.[14] However, in that particular case, a ZnS layer was identified to form directly at the buffer/absorber interface.[15] Further the CBO was *estimated* based on the measured valence band offset (VBO) and the difference in optical (bulk) band gap energies of the absorber and buffer material.[14] However, since the surface bandgap of chalcopyrite absorbers strongly differs from their bulk bandgap,[16] the actual CBO might vary significantly.

To directly access the electronic structure at the Zn(O,S)/CIS interface and ultimately shed light on the V$_{OC}$ loss of solar cells based on CuInS$_2$ absorbers grown under Cu-rich and Cu-poor conditions respectively, we use electrical characterization of respective devices in combination with ultraviolet (UPS), soft x-ray (XPS), and hard x-ray (HAXPES) photoelectron spectroscopic as well as inverse (IPES) photoelectron spectroscopic measurements of the corresponding buffer/absorber interfaces. For this, devices are fabricated with wet-chemically deposited Zn(O,S) buffer layers. The solar cells are analyzed using current-voltage (I-V) and temperature-dependent current-voltage (I-VT) measurements. The measurements reveal the presence of significant charge carrier recombination at the buffer/absorber interface in $CuInS_2$ devices realized with absorbers grown under Cu-rich conditions, but not in those based on Cu-poor absorbers. To elucidate the origin of the different recombination pathways, the energy level alignment at the respective $Zn(O,S)/CuInS_2$ interfaces is explored with particular emphasis on directly probing not only the VBO but also the CBO. For this purpose thickness series of Zn(O,S) deposited on Cu-rich and Cu-poor $CuInS_2$ absorbers, respectively, are studied using direct and inverse photoemission. We find a small positive "spike"-like CBO at the interface in both cases, excluding an unfavorable CBO as the cause of $V_{OC}$ loss in Cu-rich grown $CuInS_2$ – based devices. The observed changes in band bending upon interface formation also exclude Fermi level pinning as the main cause. This is in agreement with our recent finding of significant charge carrier recombination via acceptor-like defects in the vicinity of the buffer/absorber interface,[1] leading to a local limitation of the quasi-Fermi level splitting lowering the overall $V_{OC}$ of the device.

**EXPERIMENTAL**

**Sample preparation and handling:** $CuInS_2$ polycrystalline thin films were grown on molybdenum-coated soda-lime glass by the thermal co-evaporation process. In order to prepare $CuInS_2$ films with [Cu]/[In]>1 (Cu-rich) and [Cu]/[In]<1 (Cu-poor) as-grown stoichiometry two different process types were used: a 2-stage process to prepare Cu-rich films and 1-stage process was used to prepare Cu-poor films as explained in detail in Ref.[5] The Cu-rich absorbers were subjected to a 10 % KCN etching for 5 minutes to remove the $Cu_{2-x}S$ secondary phase and the Cu-poor films were subjected to a 5% KCN etching for 30 seconds to ensure reproducible surface conditions.[17-18] Following etching, the absorbers were rinsed with, and stored in deionized (DI) water until buffer deposition to avoid air exposure. Keeping a DI water film on top to protect the surface during transfer, these films were subsequently dipped in a wet-chemical bath for depositing Zn(O,S) buffer layer. For preparation of this bath solution 5.75 g of $ZnSO_4 \cdot 7H_2O$ and 6.09 g of $CH_4N_2S$ were dissolved separately in 60 ml deionized water each and heated at 84 °C for 10 minutes. After 10 minutes of preheating, 27 ml of $NH_4OH$ is added to the $ZnSO_4 \cdot 7H_2O$ aqueous solution followed by addition of the $CH_4N_2S$ solution and 53 ml water. The final volume of the solution was 200 ml. The $CuInS_2$ absorbers were then dipped in this solution at 84 °C. The entire recipe has been adapted and modified from a process reported by Hubert et al.[19] To prepare the Zn(O,S) thickness series, the chemical bath deposition (CBD) was stopped after 0.5, 1, 2, 4, 10, and 20 minutes, of which the latter deposition time represents the standard duration for buffer layers used in solar cells. After buffer deposition, the samples were rinsed in a 10% $NH_4OH$ aqueous solution and stored in DI water until transfer into the nitrogen-filled glove box (directly attached to the surface analysis system). To minimize air exposure the samples were placed into the load lock of the glovebox with a DI water layer on top; during pump down the samples were freeze-dried. In the glovebox, the samples were mounted on sample holders and then directly transferred into the surface analysis system for XPS measurements. Prior to UPS and IPES characterization, the samples were cleaned using a low-energy (50 eV) $Ar^+$ treatment for 240 min (bare absorbers) and 40 min (20 min CBD Zn(O,S) buffers), respectively.[20] For transfer to the SPring-8 light source, the samples were double bagged in the nitrogen atmosphere of the glovebox, with the outer bag containing desiccant. Prior to the HAXPES measurements, exposure of the samples to ambient conditions for approximately 2 hours during mounting and introduction into the measurement system was unavoidable. Due to time constraints the Cu poor absorber which underwent a 20 min Zn(O,S) CBD could not be measured at SPring-8 and hence the respective spectra are missing in this work.

**Fabrication and electrical characterization of solar cells:** Solar cells were fabricated using both Cu-rich and Cu-poor $CuInS_2$ absorbers employing a chemical bath deposited Zn(O,S) buffer layer (20 minute deposition time). Following buffer deposition, intrinsic zinc-oxide (i-ZnO) and aluminum-doped zinc-oxide (AZO) window layers were deposited using AJA magnetron sputtering unit. Finally, nickel-aluminum grids were deposited using Leybold UNIVEX e-beam evaporation unit as front contacts. To measure the elemental composition of as-grown absorbers energy-dispersive x-ray spectroscopy (EDX) at 20 kV was performed using a 20mm$^2$ area Oxford Instruments X-Max Silicon Drift Detector (SDD) installed in a Hitachi SU-70 field-emission SEM with a Schottky electron source and with a secondary electron detector.

The optical characterization was carried out by illuminating the sample using a Coherent 660 nm continuous wave diode laser. Calibrated photoluminescence (PL) measurements were used to extract the quasi-Fermi level splitting (qFLs) of the absorbers. This procedure includes two steps: first using a commercial calibration lamp a spectral correction is carried out. In the second step, an Andor CCD camera and a power meter are used to perform an intensity correction. This information is then used to tune the intensity of the laser in order to excite the sample with an intensity equivalent to one sun (*i.e.*, the same flux of photons as in an AM1.5 spectrum with energies above the band gap). The qFLs is then extracted from the PL measurements by transforming the PL flux into energy space using Planck's generalized law[21] and fitting the high energy slope.[22] A detailed description of the process can be found in Ref.[23]

For electrical characterization of the solar cells, a OAI AAA solar simulator unit with standard Xenon short-arc lamp (calibrated to 100 mW/cm$^2$ AM 1.5 using a reference Si solar cell) with a Keithley I-V source-measure-unit was used to measure the I-V characteristics of the device at room temperature (25°C). For external quantum efficiency (EQE) measurements, a Bentham EQE system was used, consisting of a halogen and a xenon lamp as light source together with a grating monochromator, a light chopper and a lock-in amplifier. To perform low-temperature electrical characterization (I-VT) a homemade setup was used. The devices were mounted inside a closed-cycle cryostat operated at $<5\times10^{-3}$ mbar. A cold mirror halogen lamp adjusted to an intensity of ~100 mW/cm$^2$ was used to illuminate the device for I-VT measurements. The equivalent calibration of the lamp was done by adjusting the lamp to sample distance and obtaining a short circuit current ($I_{sc}$) equal to the one measured under the solar simulator. To ensure accurate determination of the device temperature during characterization, a Si-diode sensor glued onto an identical glass substrate was placed beside the solar cell. For capacitance profiling same setup was used together with an inductance, capacitance, and resistance meter for frequency in the range f = 20 Hz to 2 MHz with a controlled small-signal ac voltage pulse of 30 mV root mean square and a dc bias.

**Laboratory-based photoelectron spectroscopy:** X-ray (XPS) and ultraviolet (UPS) photoelectron spectroscopy measurements were conducted using laboratory excitation sources, *i.e.* non-monochromatized Mg K$_\alpha$ (1253.56 eV, referred to as 1.3 keV in the manuscript, Specs XR 50) and He II (40.8 eV, Prevac UVS 40A2) excitation, respectively. The photoelectrons were detected by a ScientaOmicron Argus CU electron analyzer. The pass energy for the core level detail spectra measurements was set to 20 eV, resulting in an experimental energy resolution of approximately 0.9 eV for the Mg K$_\alpha$ source. For the He II measurements, the pass energy was set to 4 eV resulting in an energy resolution of approximately 0.1 eV. The binding energy (BE) of the XPS measurements was calibrated by referencing the Au 4f$_{7/2}$ peak of a clean, grounded Au foil to a binding energy of 84.00 eV. The UPS binding energy was calibrated by referencing the Fermi edge of a clean, grounded Au foil to a binding energy of 0.00 eV.

Inverse photoelectron spectroscopy (IPES) was performed in the same chamber, using a Kimball Physics Inc. EGPS-1002E electron gun with BaO coated filament and an OmniVac IPES1000 channeltron-based counter. IPES spectra were recorded in isochromat mode (*i.e.* using a bandpass filter to measure photons of constant energy, varying the energy of the incident electrons) with the detected photon energy calibrated to 6.69 eV using the Fermi edge of a clean, grounded Au foil. The kinetic energy of the electrons used for excitation was swept from 5-15eV in 0.1 eV steps with total dwell time (*i.e.*, time per energy at each scan) of 20 s/step.

The base pressure of the analysis chamber for the XPS (UPS/IPES) measurements was $5\times10^{-9}$ mbar ($6\times10^{-10}$ mbar).

**Synchrotron-based hard x-ray photoelectron spectroscopy:** Hard x-ray photoelectron spectroscopy (HAXPES) experiments were conducted at the National Institute for Materials Science (NIMS) contract beamline BL15XU of the Super Photon ring-8 GeV (SPring-8) electron storage ring.[24] The end station is equipped with a Scienta R4000 electron energy analyzer with the following geometrical setup for beamline and analyzer: horizontally polarized x-rays, with the analyzer entrance lying in the polarization plane. The x-rays hit the sample in grazing incidence and emitted electrons are detected at an angle close to 90° relative to the sample surface. Spectra were excited using a calibrated photon energy of 5.95 keV (referred to as 6 keV in the manuscript) using the Si (111) crystal of a double crystal monochromator and a Si(333) channel-cut monochromator. A pass energy of 200 eV was used for all measurements, resulting in a combined analyzer plus x-ray energy resolution of approximately 0.25 eV for all HAXPES spectra. The binding energy was calibrated by referencing the Fermi edge of a grounded clean Au foil to a binding energy of 0.00 eV. The base pressure of the setup was $< 1\times10^{-9}$ mbar.

**Curve fit analysis:** XPS and HAXPES core level spectra were fitted using linear backgrounds and Voigt profiles, keeping – if more than one species is present – the interspecies distances of one core level constant for all excitation energies and keeping the shape of the Voigt profile identical for identical core levels and excitation energies. To consider spin-orbit coupling, two Voigt profiles with a fixed distance in binding energy and a fixed intensity ratio according to $\frac{1+2(l+1/2)}{1+2(l-1/2)}$ were used.

## RESULTS AND DISCUSSION

Figure 1 (a) shows the measured I-V characteristics of solar cells under AM 1.5 illumination made from Cu-rich and Cu-poor $CuInS_2$ absorbers having an EDX-derived (*i.e.*, bulk) [Cu]/[In] ratio of 1.75 and 0.98, respectively, prior KCN etch. Thus, the seemingly high ratio for the Cu-rich absorber might not represent its true bulk composition as it is influenced by the well-known presence of the $Cu_{2-x}S$ secondary surface phase. The device I-V parameters are reported in table inset Figure 1a along with the quasi-Fermi level splitting (qFLs) of the Cu-rich and Cu-poor $CuInS_2$ absorbers measured by calibrated photoluminescence (Figure S1). As observed earlier,[5] the Cu-rich absorbers exhibit higher qFLs compared to their Cu-poor $CuInS_2$ counterparts. Between the two devices, the Cu-rich $CuInS_2$ device exhibits superior FF, $V_{OC}$ and therefore PCE value (inset in Figure 1(a)) compared to the Cu-poor counterpart. Although the $J_{SC}$ of the Cu-rich device is significantly lower than the Cu-poor devices. The $V_{OC}$ and PCE values obtained are similar to earlier reported device parameters of high-efficiency $CuInS_2$ devices based on Cu-rich and Cu-poor $CuInS_2$ absorbers.[11-13, 25] Contrary to $CuInSe_2$ (both with and without Ga), where the Cu-poor devices exhibit a higher PCE than the Cu-rich ones,[26] the Ga free Cu-poor $CuInS_2$ devices exhibit lower PCE compared to Cu-rich devices. This is because, unlike $CuInSe_2$, the Cu-rich $CuInS_2$ absorbers possess a higher optoelectronic quality than the Cu-poor absorbers (see table inset Fig. 1).[5, 23] The $J_{SC}$ of the Cu-rich devices is smaller due to a higher bandgap of ~1.51 eV compared to a bandgap ~1.48 eV of the Cu-poor absorbers ($E_g$ obtained by $\frac{d(EQE)}{dE}$ analysis[27] see Figure 1b) and a lower EQE in the longer wavelength region see Figure 1b. The device seems to suffer from a small space charge region or diffusion length that causes a drop in EQE in the longer wavelength region.[28-29] This is supported by the fact that the Cu-rich devices discussed here have a relatively high doping concentration of > $1\times10^{17}$ cm$^{-3}$ resulting in a narrow SCR compared to Cu-poor devices, for which we find a lower absorber doping concentration of ~ $1\times10^{16}$ cm$^{-3}$ (see Fig. S1c).

Despite the higher $V_{OC}$, the Cu-rich devices show a higher qFLs/e-$V_{OC}$ difference of 120 mV, whereas the Cu-poor device the value is 71 mV. A large difference between qFLs/e and $V_{OC}$ indicates losses at interfaces and contacts.[1] In optimized chalcopyrite solar cells, this difference is less than 10 mV,[30] whereas in devices with non-optimized contacts and windows but without interface recombination the difference can be 60 mV,[23] close to the difference we observe for the Cu-poor CuInS$_2$ – based device here. This suggests that the Cu-poor device might dominated by charge carrier recombination in the bulk whereas Cu-rich device is dominated by interface in agreement with the findings of Kim et al.[31] This fact is further supported by I-V measurements done at different temperatures to reveal the dominant charge carrier recombination mechanism in the devices (see Figure S2). An activation energy ($E_a$) of 1.34 (± 0.02) eV and of 1.44 (± 0.05) eV for the dominant charge carrier recombination mechanism in the Cu-rich and Cu-poor CuInS$_2$ – based device, respectively, is obtained by a linear extrapolation of the $V_{OC}$ values to 0 K (see Figure S2c).[32] For the Cu-rich CuInS$_2$ – based device, the value of $E_a$ is significantly lower than the bulk bandgap ($E_g$) of CuInS$_2$ (of 1.51 eV obtained by $\frac{d(EQE)}{dE}$ analysis, see Figure S1), in agreement with earlier works.[6, 31, 33] For the Cu-poor CuInS$_2$ device $E_a$ agrees (within the experimental uncertainty) with $E_g$. Although the confidence in $E_a$ is rather low as the device exhibits distortions in its I-V behavior at lower temperatures in an 'S shape' (see Figure S2b), which is also the reason why Cu-poor device has a rather poor FF. Despite this uncertainty, the I-V measurements combined with photoluminescence measurements confirm the presence of interface recombination in Cu-rich $CuInS_2$ – based device.

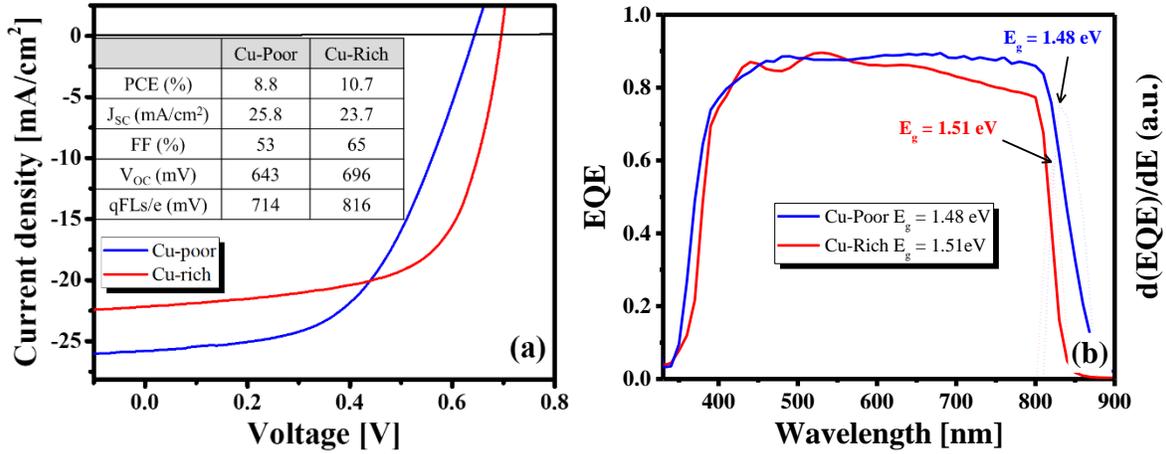

*Figure 1. (a) Experimentally measured I-V curve of Cu-rich and Cu-poor $CuInS_2$–based devices prepared with Zn(O,S) buffer layer together with derived device parameters (PCE, $J_{SC}$, FF, $V_{OC}$) and the quasi-Fermi level splitting (qFLs/e). (b) External quantum efficiency of Cu-rich (red) and Cu-poor (blue) $CuInS_2$ absorbers with Zn(O,S) layers deposited using CBD with derived $CuInS_2$ band gap energies. The energy derivative of EQE is shown in dots and the peak energy gives the bandgap of the absorber. The legend states the corresponding bulk band gap energies derived from $\frac{d(EQE)}{dE}$ analysis[27].*

The presence of a negative conduction band offset (*i.e.*, a "cliff") at the buffer/absorber interface or Fermi-level pinning at the interface are the textbook reasons for an activation energy $E_a < E_g$ in the device.[34] In the first case, $E_a$ of the dominant recombination path is equal to the interface bandgap at the buffer/absorber interface, *i.e.* $E_a = E_{g,IF} = E_g + CBO$. Whereas, in the second case, $E_a$ of the dominant recombination path is equal to the hole barrier at the buffer/absorber interface, *i.e.* $E_a = \varphi_b$. In addition to these well known effects, we recently showed, that the presence of a significantly high concentration (greater than doping concentration of the absorber) of deep defects near the CdS/CuInSe$_2$ interface (a so called p+ layer) can also cause a $E_a < E_g$ situation.[1] The qFLs in the vicinity of the interface would be significantly reduced by these deep defect states. Thus resulting in a recombination current dominated by the near interface quasi-Fermi level for electrons. All the three scenarios are depicted in Figure 2.

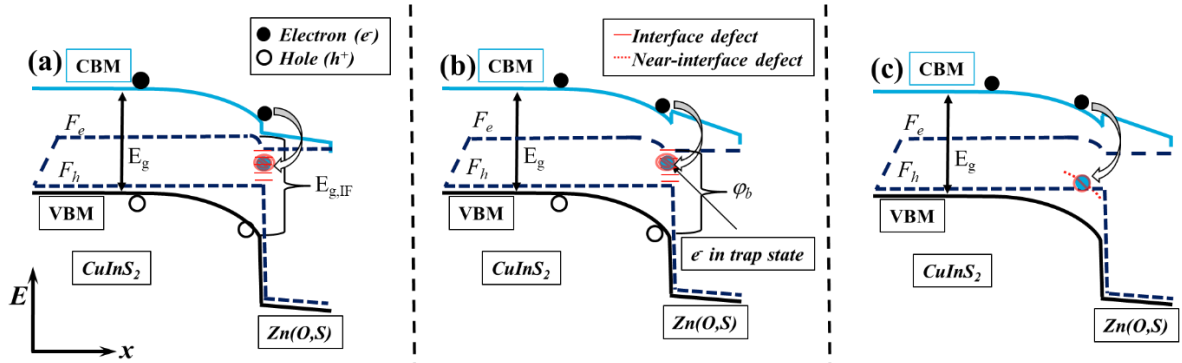

*Figure 2: Schematic of the energy level alignment of Zn(O,S)/CuInS2 interface at VOC conditions, depicting three possible scenarios (a) cliff-like CBO and (b) Fermi-level pinning due to interface defects, (c) quasi-Fermi level gradient due to near interface defects. Here $F_e$ and $F_h$ indicate the quasi-Fermi levels for electrons and holes, respectively. All of the indicated scenarios cause the activation energy ($E_a$) of the dominant charge carrier recombination process to be smaller than the absorber bandgap ($E_g$). In (a), $E_a$ equals the interface bandgap ($E_{g,IF}$) and in (b) $E_a$ is represented by the hole barrier at the buffer/absorber interface ($\varphi_b$). Where $\varphi_b$ is (nearly voltage independent in the case of Fermi level pinning, but depends on the voltage in the case of near-interface defects.[1]*

The energy level alignment (*i.e.*, whether or not a cliff-like CBO is present) at the buffer/absorber interface can be directly measured by direct and inverse photoemission measurements. Therefore, we address the Zn(O,S)/CuInS$_2$ interface by a combination of UPS, XPS, HAXPES, and IPES measurements.

To examine the energy level alignment at the buffer/absorber interface two thickness series of Zn(O,S) deposited on Cu-rich and Cu-poor CuInS$_2$ absorbers having EDX-derived (bulk) [Cu]/[In] ratios of 1.27 and 0.96, respectively, prior KCN etch. The thickness series were produced by varying the CBD time and investigated with (photon-energy dependent) direct and inverse photoelectron spectroscopy. The HAXPES (h$\nu$ = 6 keV) and XPS (h$\nu$ = 1.3 keV, Mg K$_\alpha$) survey scans (Figure S3, Figure S4) confirm the presence of all CuInS$_2$-related photoemission and Auger lines, as expected. In addition, peaks ascribed to oxygen and carbon can also be observed, which we attribute to surface adsorbates. On samples with Zn(O,S) buffers, the corresponding Zn, S, and O signals can be seen, increasing in intensity with CBD time, *i.e.* buffer layer thickness. At the same time, the CuInS$_2$ absorber lines decrease in intensity due to attenuation of the photoemission signal by the buffer. Combining the Cu 2p with the In 3d and the Cu 3p with the In 4d core levels derived by XPS and HAXPES, an inelastic mean-free path (IMFP) and therefore depth dependent [Cu]/[In] profile of the surface region can be obtained (Figure S5), displaying a similarly Cu poor surface region for both absorbers with [Cu]/[In] ratios < 0.6 for the upmost surface.

The comparison of absorber-related core level attenuation derived by XPS and HAXPES indicates that the buffer layer may not cover the (rough) CuInS$_2$ absorber homogeneously, as discussed extensively in the S.I. part "Determination of Zn(O,S) thickness". However, no absorber related signals can be observed for the more surface sensitive 1.3 keV (Mg K$_\alpha$) measurements of the 20 min Zn(O,S)/CuInS$_2$ samples (Figure S6, Figure S7). Thus, it can be concluded that in this case (*i.e.*, for the standard deposition time used to deposit the buffer for solar cells) the buffer layer is continuous and does prevent direct contact between absorber and window.

As the electronic properties of the buffer strongly depend on the S/O ratio,[35] we study the Zn(O,S) stoichiometry next. The fits of the 6 keV HAXPES Zn 3p spectra of the different samples are displayed in Figure 3. The spectra can only be reasonably fitted using (at least) two contributions: Zn$_a$ and Zn$_b$, which we tentatively assign to Zn-S and Zn-O bonds, respectively. This assignment is corroborated by the Zn$_a$ contribution having a 1:1 ratio with the ZnS-related $S_b$ component of the S 2p line as discussed in conjunction with Figure S14 and Figure S15. The calculated $Zn_a/(Zn_a+Zn_b)$ ratio directly relates to the 'x' of the $ZnO_{1-x}S_x$ composition if a 1:1 = cation:anion ratio is assumed. The value of x ranges between 0.7 and 0.9 for the Zn(O,S) thickness series and is plotted in the upper right panel of Figure 3. Referencing to previous work on the optical properties of the $ZnO_{1-x}S_x$ systems, an x-value of 0.8, which is in good agreement with our buffer composition, relates to a band gap of approximately 3.1 eV. [35-36]

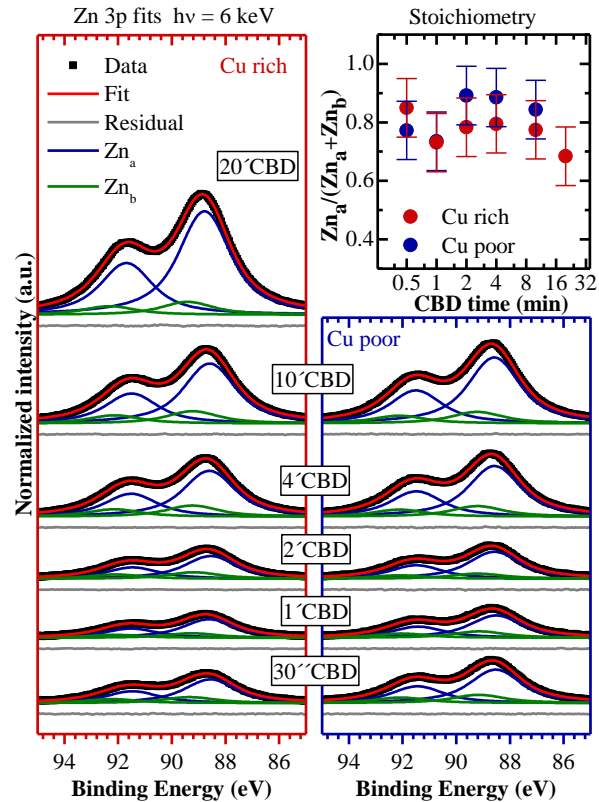

*Figure 3:* HAXPES (6 keV) spectra of the Zn 3p peak of Cu-rich (left) and Cu-poor (right) CuInS$_2$ samples with Zn(O,S) buffer layer deposited by different CBD times (from 30 s to 20 min) as indicated. Data are shown with a linear background subtracted. Fits using pairs of Voigt profiles to represent the respective doublets are displayed along the data as well as the respective residuals. The calculated $Zn_a/(Zn_a+Zn_b)$ ratio derived from the displayed fits is shown in the top right panel on a semi-log scale.

The UPS and IPES measurements on the Cu-poor and Cu-rich bare $CuInS_2$ absorbers and respective Zn(O,S) buffers prepared by 20 min CBD give information about the occupied and unoccupied density of states (DOS), respectively. The data sets are shown in Figure 4 together with linear extrapolations of the respective leading edges to determine the position of the band onsets, *i.e.,* the valence band maximum (VBM) and conduction band minimum (CBM), with respect to the Fermi level. The extrapolations of the leading edges of all UPS and bare $CuInS_2$ IPES measurements use the clearly dominant edge, as in several previous similar studies.[16, 37-38] However, our procedure for estimating the onset of the Zn(O,S) IPES spectra differs and requires more detailed explanation and justification, because in this case we propose that the low-intensity tail of the IPES spectrum, rather than the sharply increasing region of the spectrum, represents the true CBM related band edge. The explanation for this is based largely on the DOS of the Zn(O,S) conduction band, which depends strongly on the O/S stoichiometry.[35] Considering the respective binaries, calculations indicate that ZnS has a high DOS close to the CBM and thus has a very steep onset while the onset in ZnO is much more subtle, caused by a largely reduced DOS close to the CBM.[35] Furthermore, it is predicted that the CBM of ZnS is located further away from the Fermi level than that of ZnO. Since even in Zn(O,S), the Zn-anion bond lengths of the binaries are preserved, the bonds are ZnO and ZnS-like also in the alloy.[35] Thus, our measured IPES spectra of the Zn(O,S) buffer represent a superposition of these (low intensity, near Fermi level) ZnO- and (high intensity, far from Fermi level) ZnS-derived DOS, *i.e.* we expect a very shallow (ZnO related) onset followed by a very pronounced increase related to a ZnS-like DOS.[35] Thus the ZnO-related low-intensity feature can be better seen when the IPES data are plotted on a logarithmic intensity scale. Extrapolation of the logarithmic plot of the main Zn(O,S) conduction band edge (Figure S16) delivers very similar CBM results to what is shown in Figure 4 using the low-intensity tail of the spectra, justifying the evaluation approach for the CBM of Zn(O,S).

Combining the VBM and CBM results, which are also displayed in Figure 4, gives surface bandgaps of 2.02 (± 0.14) eV and 2.00 (± 0.14) eV for the Cu-rich and Cu-poor CuInS$_2$, respectively. This is significantly larger than the respective bulk bandgaps of 1.51 eV and 1.48 eV (see Figure S1d), but are however, comparable to previously determined surface bandgaps of CuInS$_2$ absorbers.[39] Such a bandgap widening towards the surface has also been previously observed for selenide chalcopyrite absorbers,[16] and might be related to the observed Cu depletion towards the surface in both absorbers (Figure S4).[39] The bandgap widening is expected to be driven by a VBM shift away from the Fermi level towards the surface, as the p-d repulsion of the Cu 3d and S 3p derived states forming the valence band edge is reduced for lower Cu contents.[40] Such a shift can indeed be found in our samples, as can be seen when comparing the less surface sensitive HAXPES and very surface sensitive UPS derived VBM values (Figure S17). However, this shift might – in addition to the band gap widening towards the surface – also in part be caused by a downward band bending towards the surface. More specific, especially the difference in UPS derived VBM values between Cu-poor and Cu-rich absorber can be assigned to increased surface band bending for the Cu-rich CuInS$_2$, as the derived surface bandgaps are very similar in both cases.

The surface band gaps of the Zn(O,S) buffers on the two absorber types are found to be 3.06 (± 0.22) eV (Cu-rich CuInS$_2$) and 3.19 (± 0.22) eV (Cu-poor CuInS$_2$), in agreement with the 3.1 eV band gap expected based on the [O]/[S] stoichiometry, as discussed above.

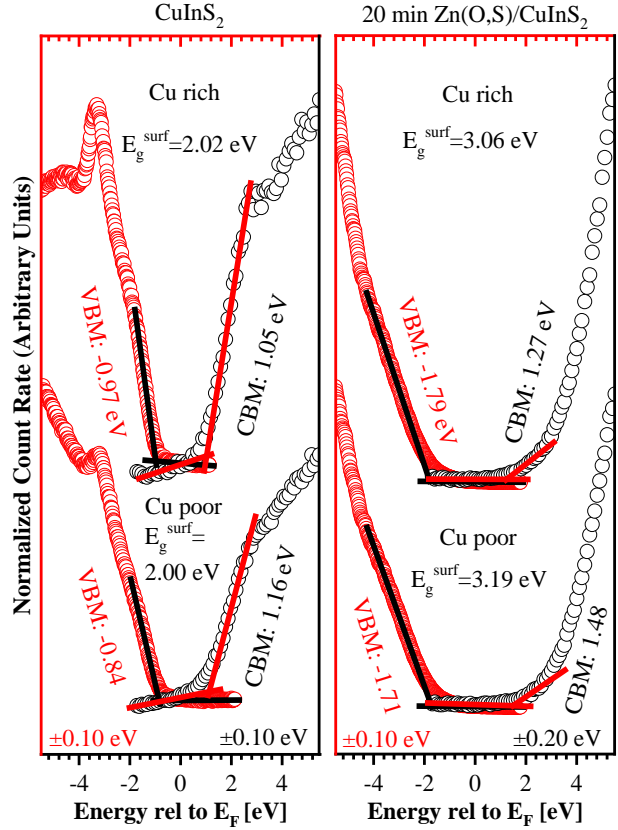

*Figure 4:* UPS (red) and IPES (black) spectra of Cu-rich (top) and Cu-poor (bottom) CuInS$_2$ absorbers without (a) and with a 20 min CBD Zn(O,S) buffer (b). All samples were treated with 50 eV Ar+ ions prior measurements. IPES and UPS spectra are shown on a common energy axis relative to the Fermi level. Solid red (IPES) and black (UPS) lines display the linear extrapolation used for determination of the valence band maximum (VBM) and conduction band minimum (CBM) with respect to the Fermi level ($E_F$). The VBM and CBM values as well as the derived electronic surface band gap ($E_g^{surf}$ = CBM-VBM) are indicated. The experimental uncertainty of the VBM and CBM values of the CuInS$_2$ absorbers and of the VBM value of the Zn(O,S)/CuInS$_2$ samples is ± 0.1 eV. For the CBM of the buffer layer we attribute an uncertainty of ± 0.2 eV, see text for explanation.

The VBM and CBM values of the bare Cu-poor and Cu-rich $CuInS_2$ absorber and the 20 min Zn(O,S) buffer depicted in Figure 4 already give a coarse (zeroth order) approximation of the interfacial energy level alignment. In both cases, *i.e.* independent of Cu-content of the absorber, we find a negative VBO (*i.e.,* the VBM of the absorber is above that of the buffer) and a positive CBO (*i.e.,* the CBM of the absorber is below that of the buffer) at the buffer/absorber interface. However, the effect of junction formation on the energy levels must also be included in this consideration. This is done by calculating the interface-induced band bending (IIBB).[37] The IIBB can directly be derived from the change of binding energies of $CuInS_2$ and Zn(O,S) related core levels upon buffer deposition. For this, the corresponding XPS data of buffer/absorber samples of the Zn(O,S) thickness series for which both layers could be probed was evaluated. The IIBB was obtained by combining the average of the buffer deposition-induced Cu 2p and In 3d core level shifts relative to the bare $CuInS_2$ absorber with the shifts of the buffer Zn 2p core level relative to the 20 min CBD Zn(O,S) buffer. The resulting IIBB values for all samples with a buffer layer deposition time of one minute or longer are displayed in Figure 5 along with the resulting averages of 0.14 eV (Cu-rich $CuInS_2$ interface) and 0.22 eV (Cu-poor $CuInS_2$ interface) and respective standard deviations of ±0.08 eV (Cu-rich) and ±0.02 eV (Cu-poor). The absolute binding energy values of all respective core levels can be found in Figure S6, Figure S7, Table S1, and Table S2. As the formation of a p-n junction naturally would lead to a downward band bending in the upper region of the p-type $CuInS_2$ absorber towards the n part of the junction, the observed shift of the Cu 2p and In 3d photoemission lines with increasing buffer layer thickness to lower BE, displayed in Figure S18, seems to be surprising. However, the shift to lower absorber core level BE can also be interpreted as a reduction of the preexisting downward band bending at the $CuInS_2$ surface due to buffer deposition induced passivation of charged defects at the surface. Furthermore, the observed shift, in particular on the Cu-rich sample, indicates that Fermi level pinning is not the main factor that limits $V_{OC}$ and/or causes $E_a<E_g$.

The exact valence- and conduction band offsets (VBO, CBO) can then be calculated using IIBB together with the previously determined VBM and CBM values using the following formulae:[41]

$$VBO = VBM_b - VBM_a - IIBB \qquad (3)$$

$$CBO = CBM_b - CBM_a - IIBB \qquad (4)$$

Where IIBB is the average value depicted as dashed line in Figure 5 and the subscripts "a" and "b" stand for "absorber" and "buffer", respectively. Including the IIBB the CBO [VBO] can be determined to +0.08 (±0.24) eV [-0.96 (±0.16) eV] and +0.10 (±0.22) eV [-1.09 (±0.14) eV] for the buffer/absorber interface using the Cu-rich and Cu-poor CuInS$_2$ samples, respectively.

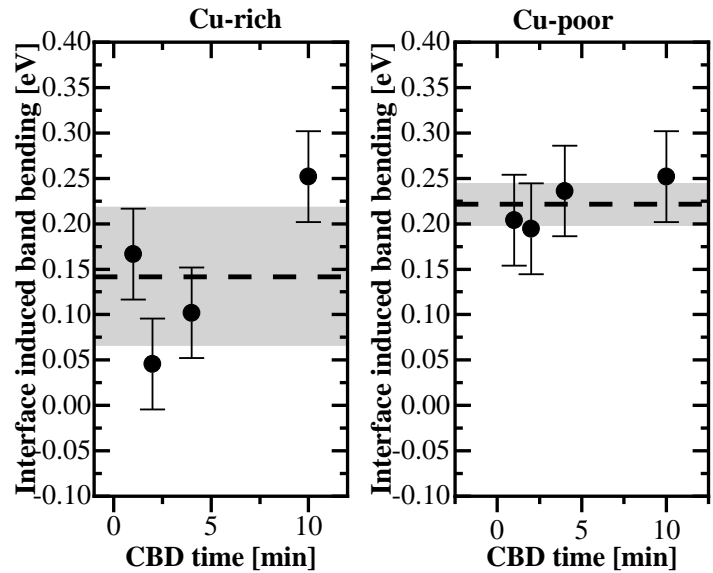

*Figure 5: Interface induced band bending (IIBB) of the Zn(O,S)/CuInS$_2$ samples using Cu-rich (a) and Cu-poor (b) absorbers. The black dashed lines give the averages of the respective sample and the grey area visualizes the standard deviation as a measure of confidence. IIBB values were calculated by combining buffer and absorber core*

*level shifts, which were determined from Cu 2p and In 3d peaks at the absorber side and from Zn 2p for the buffer. All used core level binding energies are given in Table S1 and Table S2.*

The resulting picture of the energy level alignment at the Zn(O,S)/CuInS$_2$ interface for the Cu-rich and Cu-poor absorbers is schematically shown in Figure 6. We find a significant cliff-like VBO and a small spike-type offset in the CB at the buffer/absorber interface in both cases, which is the ideal energy level alignment for this configuration.[42-43] Based on this finding alone, the studied CuInS$_2$ – based solar cells should be able to achieve high efficiencies, in the Cu-rich as well as in the Cu-poor case. The *positive* CBO does not explain the I-VT derived $E_a < E_g$ situation, which was suggested to be indicative for the presence of charge carrier recombination at the interface as the dominant (device limiting) mechanism. The observed IIBB further excludes Fermi level pinning as the main cause of an interface dominated recombination path. This leaves the presence of a near-interface defective layer as possible explanation for the experimental finding of $E_a < E_g$. Thus we can conclude, that, very similar to Cu-rich selenide,[44] the main problem of the Cu-rich sulfide is a defective near-interface layer,[1] likely caused by the necessary etching process.

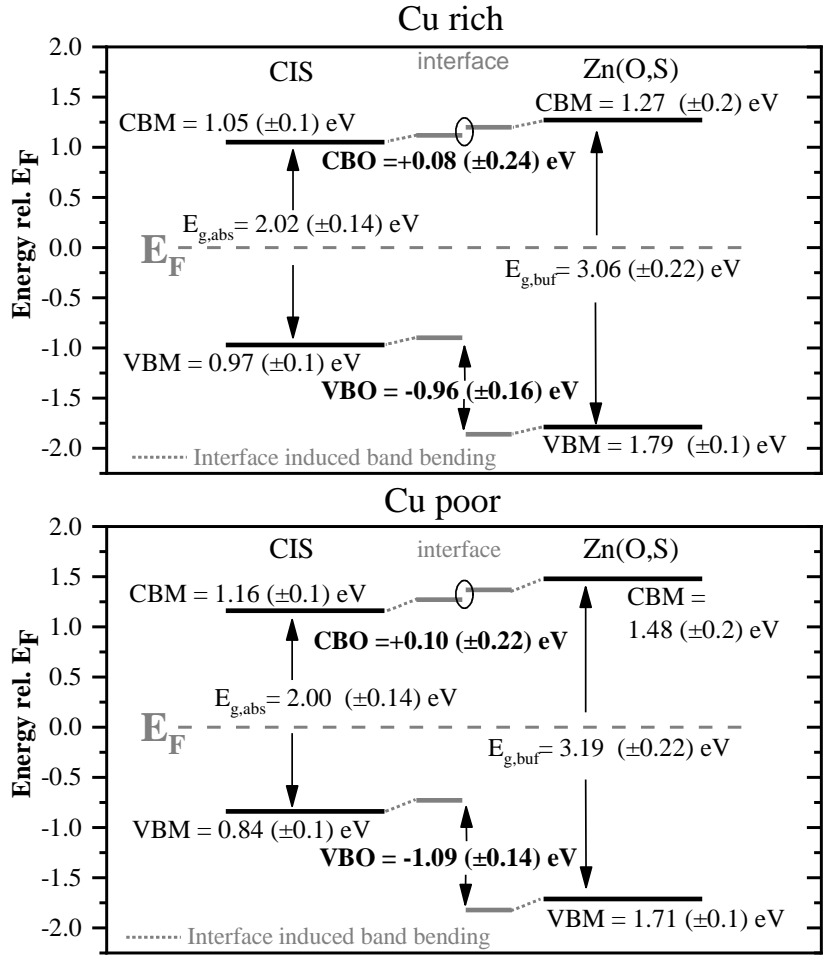

*Figure 6:* *Illustration of the energy level alignment at the Zn(O,S)/CuInS$_2$ interface when using Cu-rich (top) and Cu-poor (bottom) absorbers. The VBM and CBM values are displayed on the left (CuInS$_2$) and right (Zn(O,S)) side. The VB and CB band offsets considering IIBB are displayed in the center.*

**CONCLUSION**

We present Cu-rich and Cu-poor CuInS$_2$ absorbers and their solar cells with a Zn(O,S) buffer. As observed previously,[5, 31] we find that the performance of the Cu-rich device is limited by interface recombination. We use (direct and inverse) photoelectron spectroscopy on the absorbers and on the absorbers covered successively with the Zn(O,S) buffer to investigate differences between the Cu-rich and the Cu-poor absorber and to determine the energy level alignment at the buffer/absorber interface. We find that both absorbers have a Cu-deficient surface and exhibit a surface bandgap that is larger than the bulk bandgap. The conduction band offset between buffer and absorber in both cases is characterized by a small spike-like offset of about 0.1 eV. This excludes an unfavorable energy level alignment as the source for the dominating charge carrier interface recombination in the case of Cu-rich absorbers. Furthermore, in both cases we observe changes in the band bending during interface formation, indicating that also Fermi level pinning is not the main reason for interface recombination. This leaves the presence of near-interface defects, as the likely reason for the performance limitation of Cu-rich $CuInS_2$ based solar cells, as was recently confirmed for related $CuInSe_2$ absorbers.[1, 44]

**Acknowledgements**

MS acknowledges this research was funded in whole, or in part, by the Luxembourg National Research Fund (FNR), in the framework of the MASSENA project (grant reference [PRIDE 15/10935404]) and the CORRKEST project (grant reference [C15/MS/10386094]. For the purpose of open access, the author has applied for a Creative Commons Attributions 4.0 International (CC BY 4.0) license to any Author Accepted Manuscript version arising from this submission.

JB acknowledges support from the Graduate School Materials for Solar Energy Conversion (MatSEC) as part of Dahlem Research School. The HAXPES measurements at SPring-8 were performed under an approval of NIMS Synchrotron X-ray Station (Proposal No. 2018A4908) and was supported by NIMS microstructural characterization platform as a program of "Nanotechnology Platform" (project No. 12024046) of the Ministry of Education, Culture, Sports, Science and Technology (MEXT), Japan.

# Supporting Information

**Origin of interface limitation in CuInS$_2$ – based solar cells**

Mohit Sood*[1], Jakob Bombsch*[2], Alberto Lomuscio[1], Sudhanshu Shukla[1], Alberto Lomuscio[1], Claudia Hartmann[2], Johannes Frisch[2,3], Wolfgang Bremsteller[2,3], Shigenori Ueda[4,5,6], Regan G. Wilks[2,3], Marcus Bär[2,3,7,8], Susanne Siebentritt[1]
*equal contributions
(E-mail addresses: mohit.sood@uni.lu, jakob.bombsch@helmholtz-berlin.de)
[1]*Laboratory for Photovoltaics, Physics and Materials Science Research Unit, University of Luxembourg, Belvaux, L-4422, Luxembourg*
[2]*Department Interface Design, Helmholtz-Zentrum Berlin für Materialien und Energie GmbH (HZB), 12489 Berlin, Germany*
[3]*Energy Materials In-situ Laboratory Berlin, HZB, 12489 Berlin, Germany*
[4]*Department X-ray Spectroscopy at Interfaces of Thin Films, Helmholtz Institute for Renewable Energy (HI ERN), 12489 Berlin, Germany*
[5]*Department of Chemistry and Pharmacy, Friedrich-Alexander Universität Erlangen-Nürnberg (FAU), 91058 Erlangen, Germany*

**Calibrated photoluminescence measurements:**

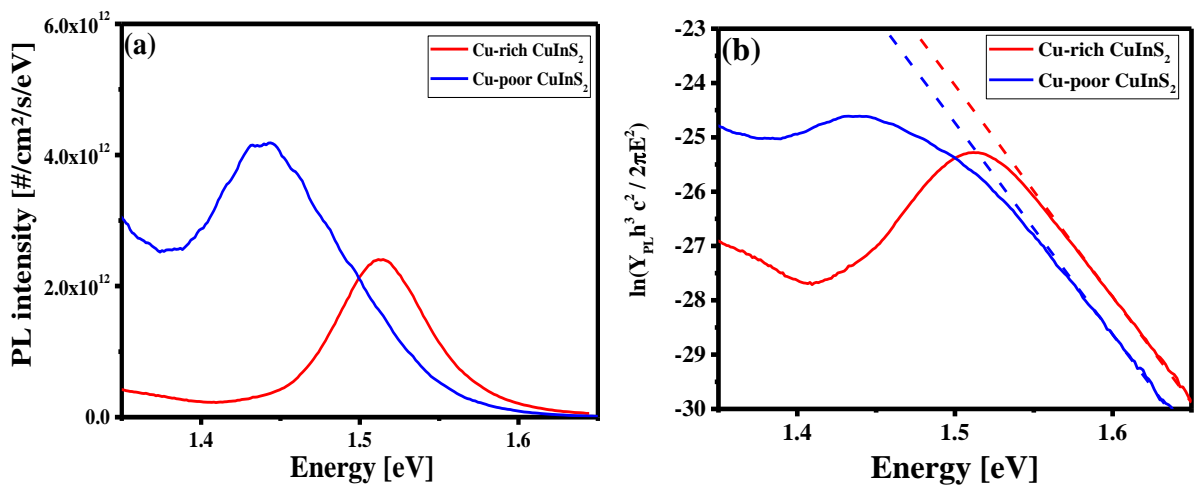

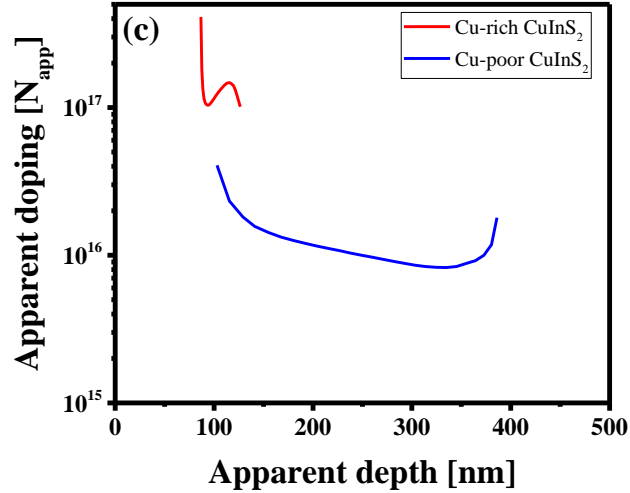

*Figure S1:* (a) Exemplary intensity-calibrated PL spectra of Cu-rich and Cu-poor CuInS$_2$ absorbers used to fabricate devices. The qFLs values are obtained using (b) transformed PL spectra of these absorbers using Planck's generalized law as explained in Ref[22-23]. The laser power for the measurement was set to 15 sun intensity. Due to poor PL signal from the sample, the qFLs at one sun was extrapolated by measuring qFLs at 4, 7, 10 and 15 suns. (c) Apparent doping profile versus apparent depth for Cu-poor and Cu-rich CuInS$_2$ solar cell.

**Activation energy for dominant charge carrier recombination and band gap energy derived from I-VT and external quantum efficiency measurements, respectively:**

Fig. S2a and b display the I-V curves of Cu-rich and Cu-poor CuInS$_2$ measured at different temperatures. For Cu-poor devices, I-V curves show distortion at lower temperatures Fig. S2b, therefore, to get an estimate of activation energy the measurements were done in smaller temperature steps. The activation energy of the dominant charge carrier recombination path in the device was obtained using the following equation:[28]

$$V_{OC,ex} \approx \frac{E_a}{e} - \frac{AkT}{e}\left(\frac{J_{00}}{J_{sc}}\right)$$

where e is the charge of an electron, E$_a$ is the activation energy of the dominant charge carrier recombination path, A is the ideality factor of the device, k is the Boltzmann constant, T is the temperature, J$_{00}$ is a weakly temperature-dependent term, and J$_{sc}$ is the short-circuit current (here it is assumed that short-circuit current is equal to photocurrent, J$_{ph}$). Under the condition that the A and J$_{ph}$ term remains weakly temperature dependent, the extrapolation of V$_{OC,ex}$ to 0 K gives the value of E$_a$.

Fig. S2d displays the ideality factor of Cu-rich CuInS$_2$ device plotted as a function of temperature obtained using the I-V fit routine[45] by fitting the I-V curves in Fig. S2a with the one diode equation using a Python script developed at University of Luxembourg. The detailed procedure for fitting can be found in Ref[46]. The ideality factor is not fully temperature independent, however, in a certain region 302-270 K it is only weakly temperature dependent. Hence, for deriving E$_a$ the linear extrapolation of the V$_{oc}$ values to 0 K was done in this region.

No such analysis could be done for Cu-poor devices, as the I-V curves show significant distortion (Fig. S2b), making it difficult to reliably obtain ideality factors at different temperatures. At higher temperatures the 'S shape' is reduced, suggesting the presence of a thermionic barrier as the cause of this I-V curve distortion.[28] Nonetheless, a fit in high temperature range was made where the device behaves a like a standard p-n diode.

The band gap energies for the Cu-poor and Cu-rich CuInS$_2$ is obtained from derivative of respective external quantum efficiency measurements (Fig. 1b). The lower bandgap and the weak absorption onset edge in Cu-poor CuInS$_2$ device might be due to the presence of band tail states,[47] which have been suggested to originate from structural disorder and high defect density in case of Cu-poor material.[26, 48] It must be noted that both Cu-rich and Cu-poor devices degrade with time, with Cu-poor devices degrading significantly faster (efficiencies reach less than 2 % within a few weeks) than Cu-rich devices.

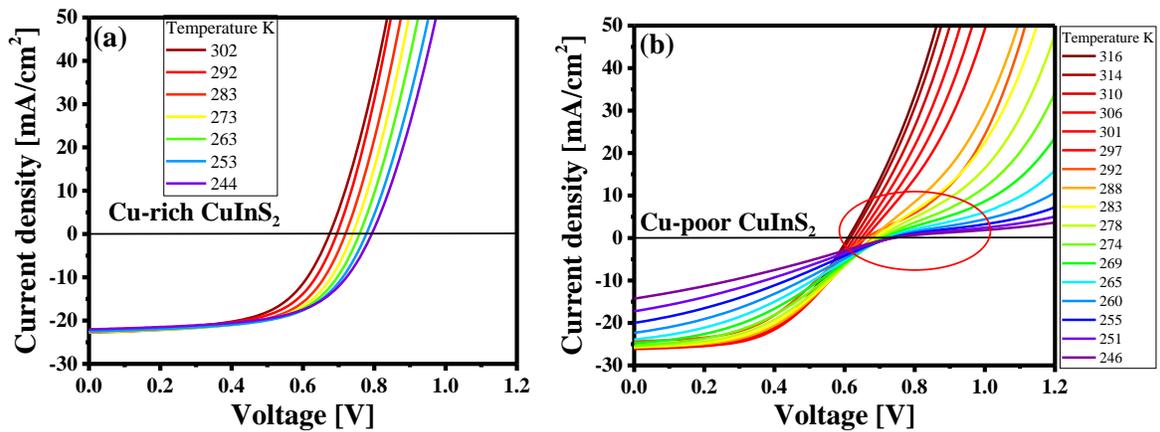

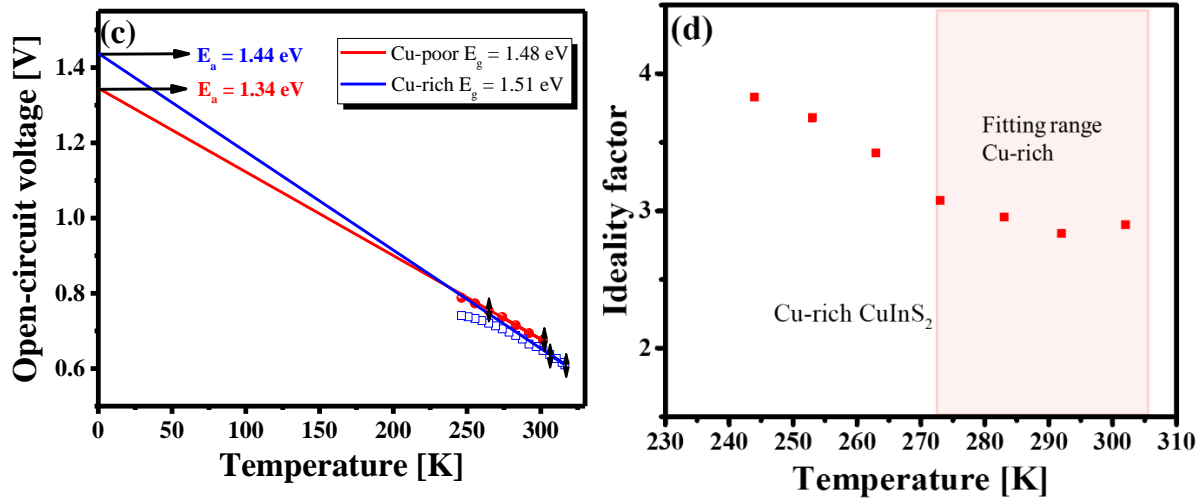

*Figure S2:* I-V curves of (a) Cu-rich CuInS$_2$ devices (b) Cu-poor CuInS$_2$ devices measured at different temperatures. (c) Open-circuit voltage ($V_{OC}$) of the devices plotted as a function of temperature. The linear extrapolation to derive the activation energy of the dominant charge carrier recombination mechanism [$E_a$ = $V_{OC}$ (0K)] is indicated together with the derived value for the Cu-rich device. The arrow bars show the fit range used to determine the $E_a$. The legend states the corresponding bulk band gap energies derived from $\frac{d(EQE)}{dE}$ analysis (see Figure 1b). (d) Ideality factor of Cu-rich CuInS$_2$ plotted as a function of temperature.

**HAXPES and XPS survey spectra:**

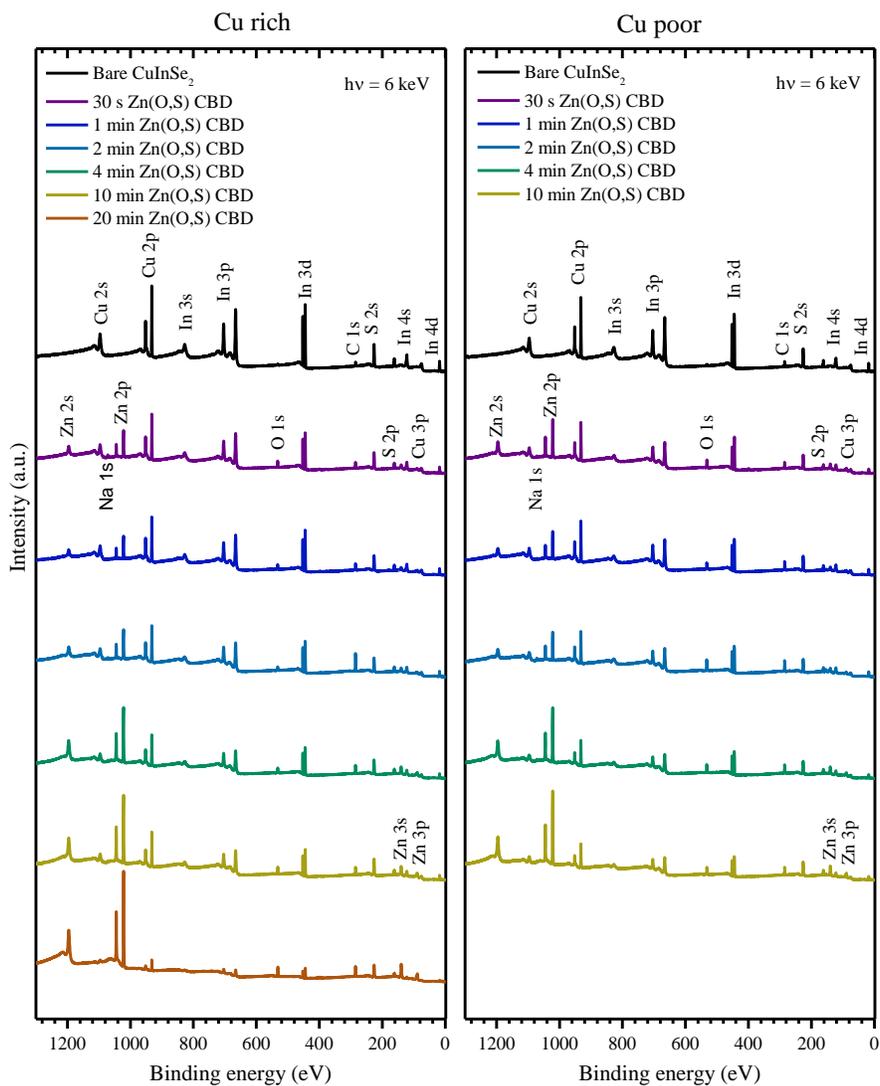

*Figure S3:* HAXPES (6 keV) survey spectra of Cu rich (left) and Cu poor (right) $CuInS_2$ absorbers with Zn(O,S) layers deposited using CBD for different durations (30 s to 20 min). Spectra are vertically offset for clarity.

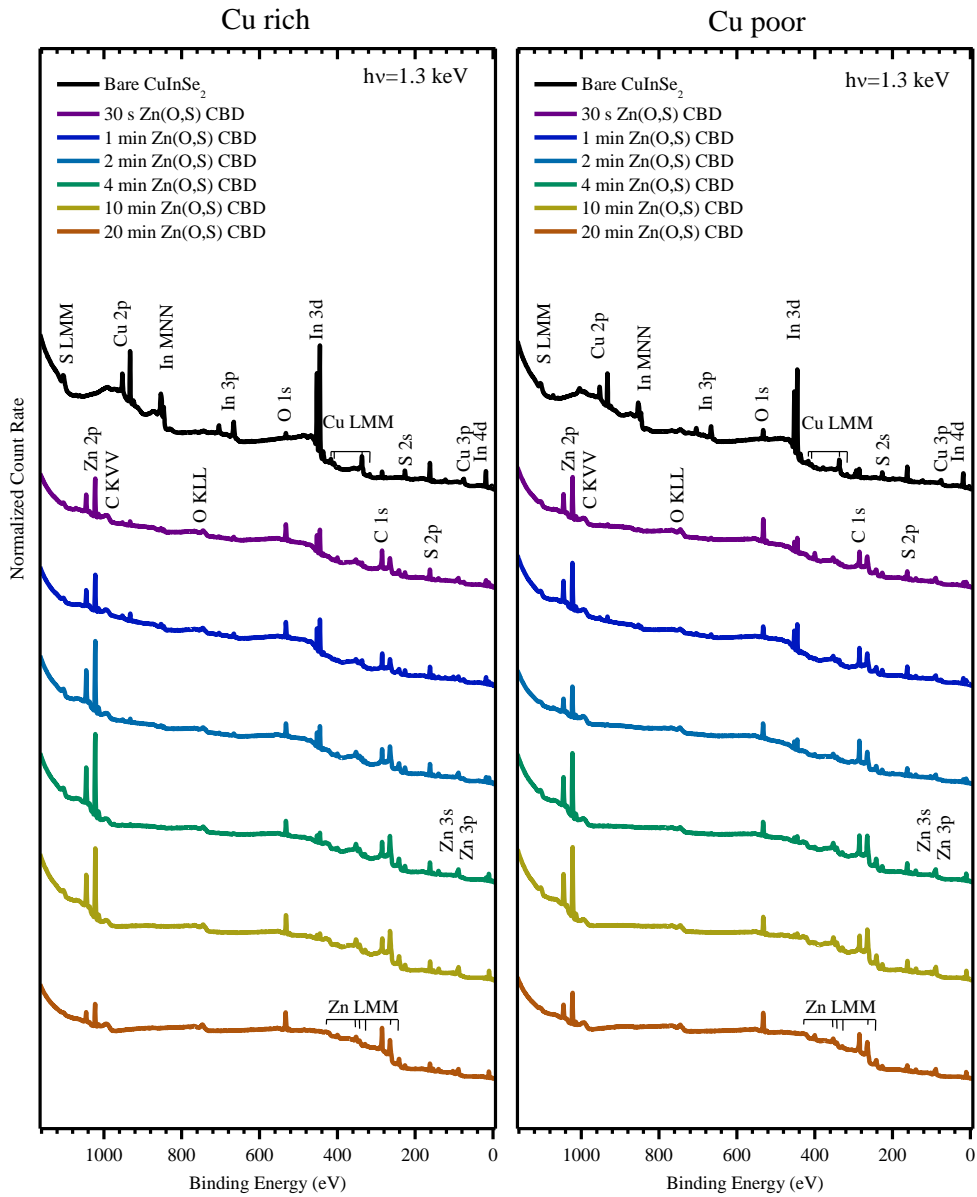

*Figure S4:* XPS (1.3 keV, Mg K$\alpha$) survey spectra of Cu rich (left) and Cu poor (right) CuInS$_2$ absorbers with Zn(O,S) layers deposited using CBD for different durations (30 s to 20 min). Spectra are vertically offset for clarity.

**CuInS$_2$ surface stoichiometry:**

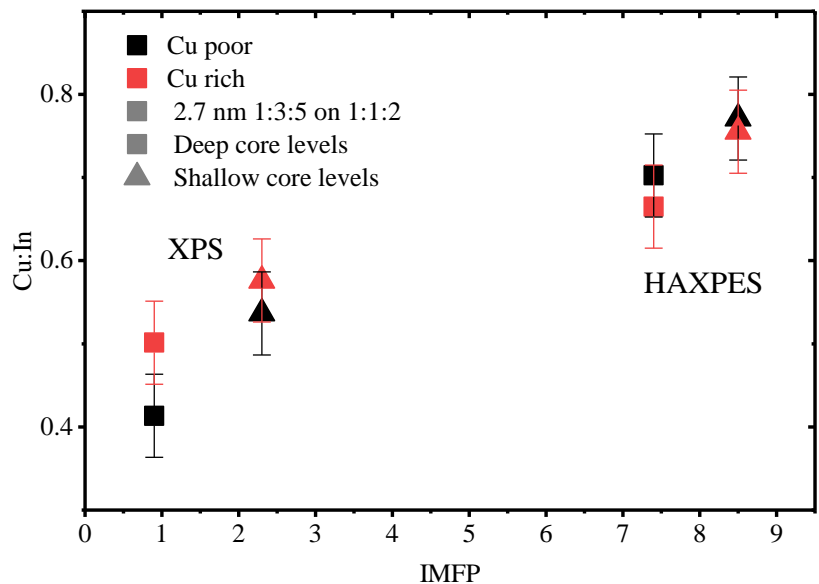

***Figure S5:*** *[Cu]/[In] ratio depth profile of the Cu rich and Cu poor CuInS$_2$ absorbers. Values are obtained by comparing deep core levels (Cu 2p and In 3d) and shallow core levels (Cu 3p and In 4d) obtained with more surface sensitive laboratory based XPS measurements and less surface sensitive HAXPES measurements.*

**XPS and HAXPES detail spectra corresponding to Cu, In, and Zn:**

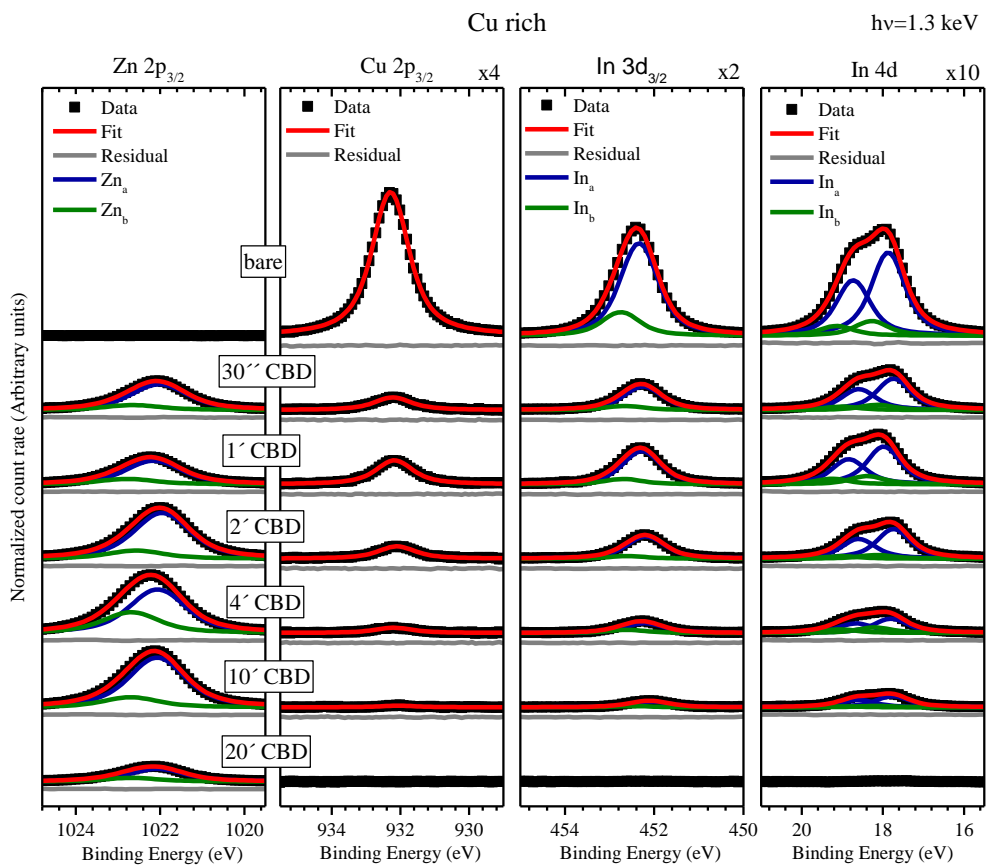

*Figure S6:* XPS (1.3 keV, Mg Kα) spectra of Cu rich CuInS$_2$ absorber with Zn(O,S) layers deposited using CBD for different durations (from 30 s to 20 min). Zn 3p$_{3/2}$, Cu 2p$_{3/2}$, In 3d$_{3/2}$, and In 4d peaks are displayed. Data are shown with a linear background subtracted. Respective fits using Voigt profiles or pairs of Voigt profiles to represent the respective doublets, are displayed along the data as well as the respective residuals. The In 3d$_{3/2}$, In 4d, and Cu 2p spectra are magnified by a factor given at the top right of the respective graphs for better visibility. Spectra are vertically offset for clarity, as are residuals.

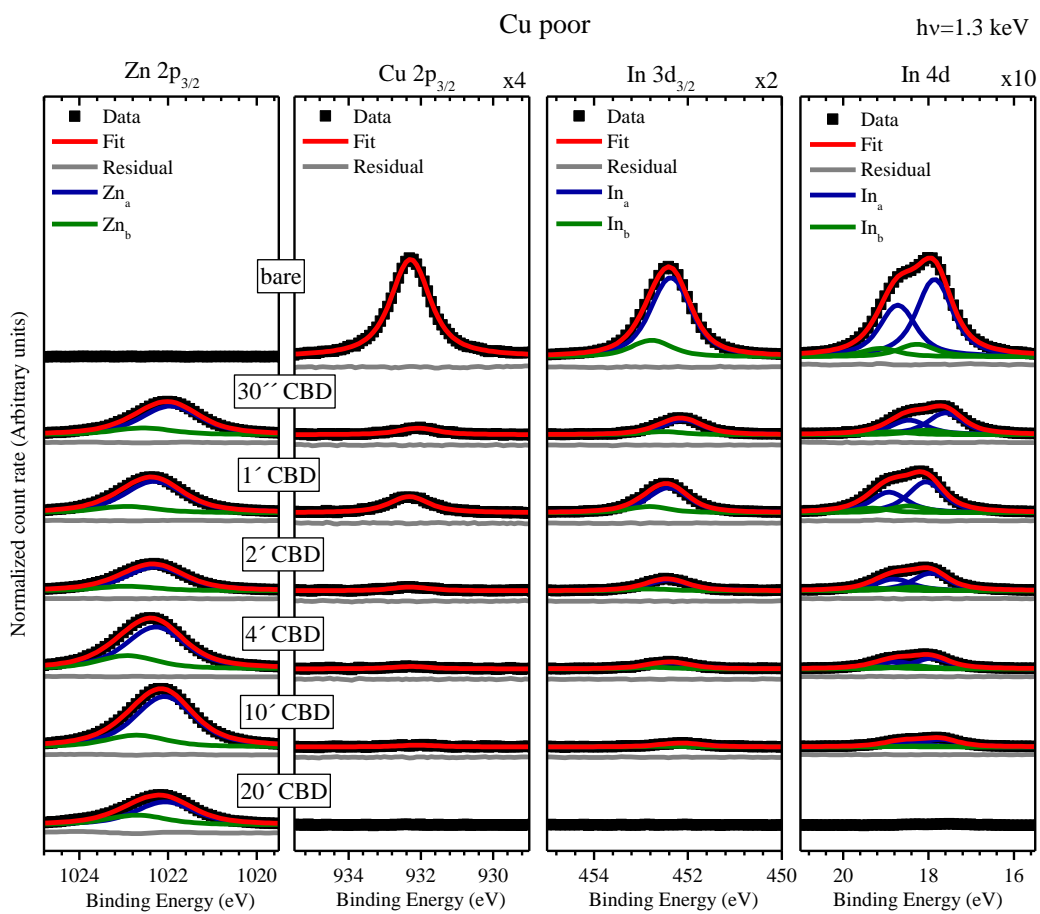

**Figure S7:** XPS (1.3 keV, Mg Kα) spectra of Cu poor CuInS$_2$ absorber with Zn(O,S) layers deposited using CBD for different durations (from 30 s to 20 min). Zn 3p$_{3/2}$, Cu 2p$_{3/2}$, In 3d$_{3/2}$, and In 4d peaks are displayed. Data are shown with a linear background subtracted. Respective fits using Voigt profiles or pairs of Voigt profiles to represent the respective doublets, are displayed along the data as well as the respective residuals. The In 3d$_{3/2}$, In 4d, and Cu 2p spectra are magnified by a factor given at the top right of the respective graphs for better visibility. Spectra are vertically offset for clarity, as are residuals.

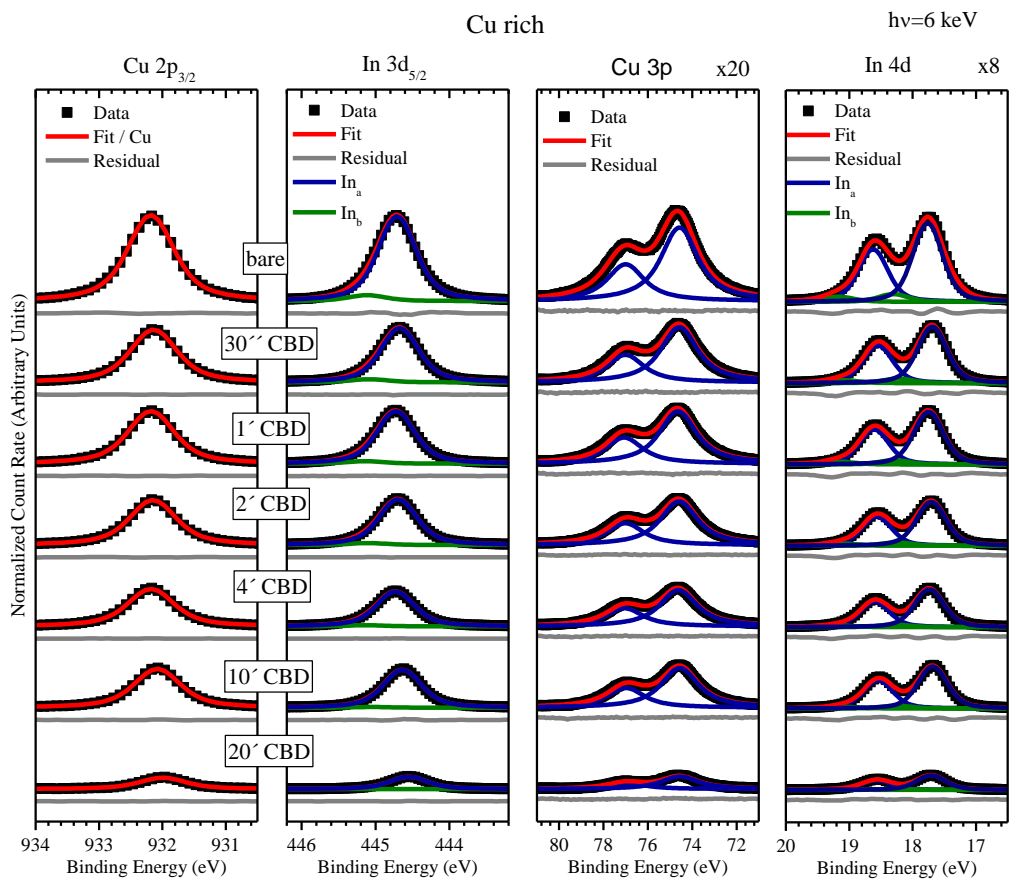

*Figure S8:* HAXPES (6 keV) spectra of Cu rich CuInS$_2$ absorbers with Zn(O,S) layers deposited using CBD for different durations (from 30 s to 20 min). Cu 2p$_{3/2}$, In 3d$_{5/2}$, Cu 3p with the In 4p background subtracted (see "In 4p-related background correction for Cu 3p HAXPES data" part) (c), and In 4d (d) peaks are displayed. Data are shown with a linear background subtracted. Respective fits using Voigt profiles or pairs of Voigt profiles to represent the respective doublets, are displayed along the data as well as the respective residuals. The In 4d and Cu 3p spectra are magnified by a factor given at the top right of the respective panels for better visibility. Spectra are vertically offset for clarity, as are residuals.

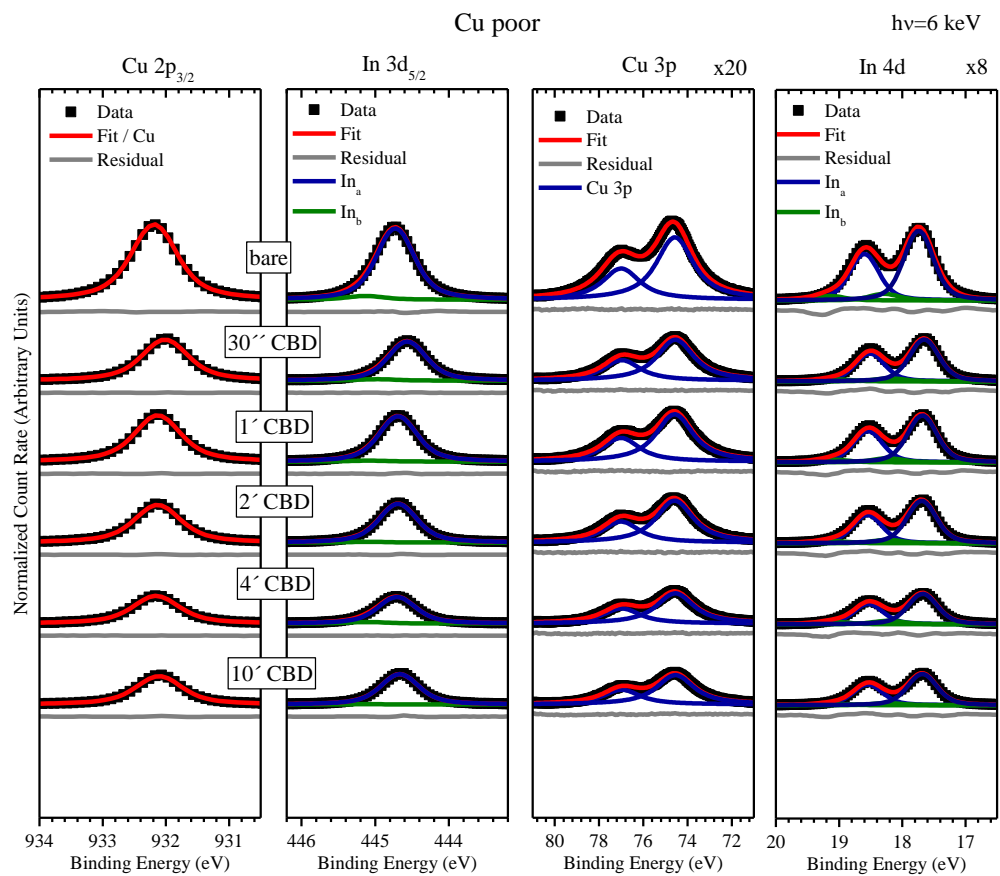

*Figure S9:* HAXPES (6 keV) spectra of Cu poor CuInS$_2$ absorbers with Zn(O,S) layers deposited using CBD for different durations (from 30 s to 20 min). In 3d$_{5/2}$ (a), In 4d (b), Cu 2p$_{3/2}$ (c), and Cu 3p (d) peaks are displayed. Data are shown with a linear background subtracted. Respective fits using Voigt profiles or pairs of Voigt profiles to represent the respective doublets, are displayed along the data as well as the respective residuals. The In 4d and Cu 3p spectra are magnified by a factor given at the top right of the respective graphs for better visibility. Spectra are vertically offset for clarity, as are residuals.

**In 4p-related background correction for Cu 3p HAXPES data:**

As the Cu 3p peak strongly overlaps with the broad In 4p peak, a removal of the In contribution is necessary prior quantification. To do so, an $In_2Se_3$ spectrum (blue lines in Figures S10 and S11) was used as reference. It was shifted, scaled (such that the In 4d line overlaps), and subtracted from the different buffer/absorber spectra (black lines in Figures S10 and S11). The resulting difference spectra are shown as red lines in Figures S10 and S11.

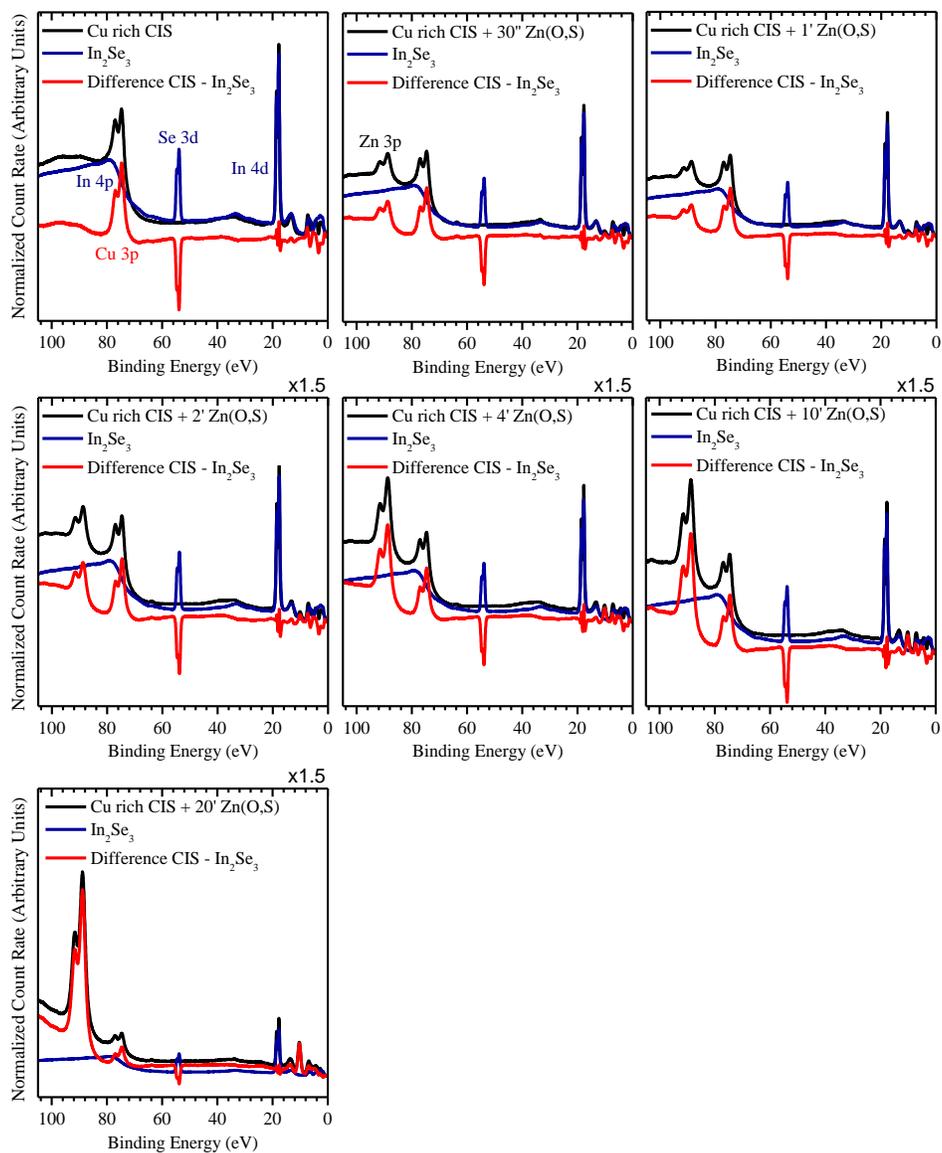

*Figure S10:* 6 keV HAXPES spectra of the shallow core level region of the buffer / Cu rich $CuInS_2$ (CIS) absorber samples are shown from top to bottom with increasing Zn(O,S) CBD duration (from 30 s to 20 min).

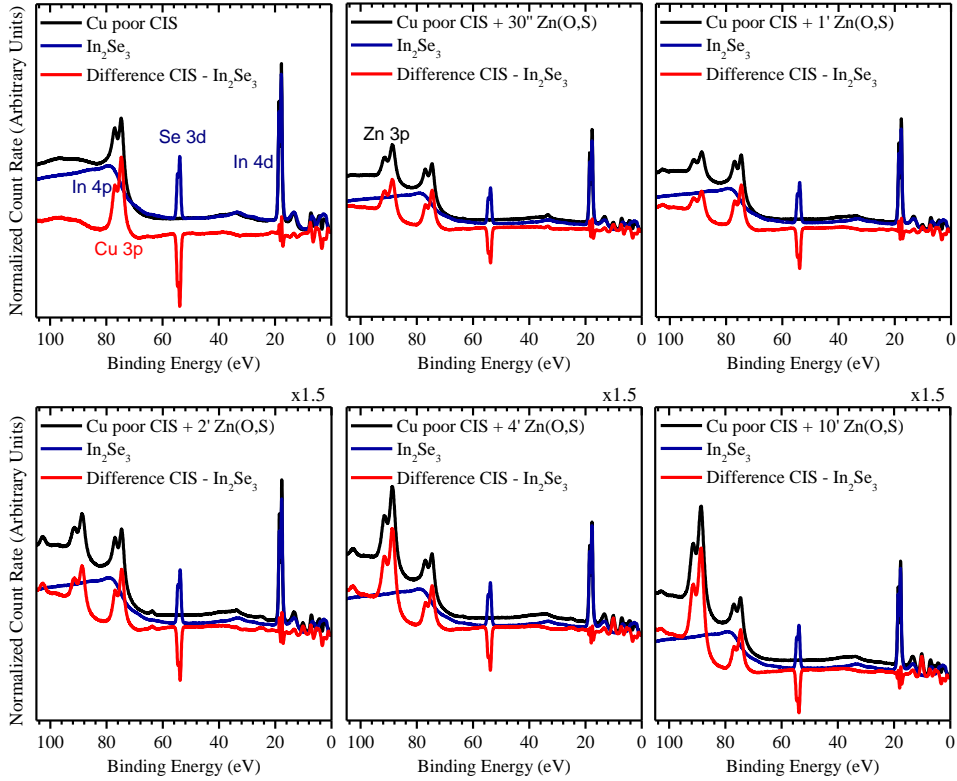

***Figure S11:*** *6 keV HAXPES spectra of the shallow core level region of the buffer / Cu poor CuInS$_2$ (CIS) absorber samples are shown from top to bottom with increasing Zn(O,S) CBD duration (from 30 s to 20 min).*

**Determination of Zn(O,S) thickness:**

The attenuation of the Cu and In photoemission lines was used to estimate the buffer thickness d. Assuming that a homogeneous buffer is completely covering the CuInS$_2$ absorber, we can use the Beer-Lambert law

$$d = IMFP * \ln\left(\frac{I_0}{I_d}\right) \quad\quad\quad (1)$$

where I$_0$ is the intensity of a core level spectrum measured from the bare absorber and I$_d$ the corresponding intensity measured from an absorber capped with a buffer of thickness d.

The inelastic mean free path (IMFP) is dependent on the attenuating material and the kinetic energy of the photoelectrons, and is determined using the QUASES IMFP TPP2M code assuming the buffer to be a 1:1 combination of ZnO and ZnS.[49-50] The core level spectra measured with 1.3 keV and 6 keV and used in these calculations can be found in **Figure S6, Figure S7, Figure S8, Figure S9, Figure S10, and Figure S11.** The average thickness values displayed in **Figure S12a** were obtained from the Cu 2p, In 3d, Cu 3p, and In 4d core levels in case of the 6 keV measurements and from the Cu 2p, In 3d, and In 4d measurements in the case of the 1.3 keV measurements (the values calculated for the individual core levels are displayed in **Figure S13)**.

This approach produces results suggesting a buffer thicknesses of about 16 nm for the thickest layer. Also, the calculated buffer thicknesses vary depending on the used excitation energy, which suggests either that there is a systematic error in the IMFP values at different photoelectron energies, or that the assumption of a homogeneous buffer thickness is incorrect. To demonstrate the effect of an inhomogeneous buffer layer, we introduce a model of a stepped surface where the buffer in the "valleys" is thicker than the detection depth of our HAXPES measurements (<30 nm). If we assume that 30% of the surfaces are covered with such buffer-filled valleys and respectively altering Formula (1) to

$$d = IMFP * \left(\ln\left(\frac{I_0}{I_d}\right) + \ln(0.7)\right) \qquad (2)$$

Using this model, the remaining 70% of the buffer is responsible for signal attenuation and the result is a good agreement between the 1.3 keV- and 6 keV-excited data, as shown in **Figure S12b**. This result demonstrates that the thicknesses derived from excitation energy-dependent photoemission data can be brought into agreement by assuming an inhomogeneous buffer thickness, however it is clear that there are any number of specific structures that will produce this effect (i.e., the result does not suggest that the stepped structure is physically present).

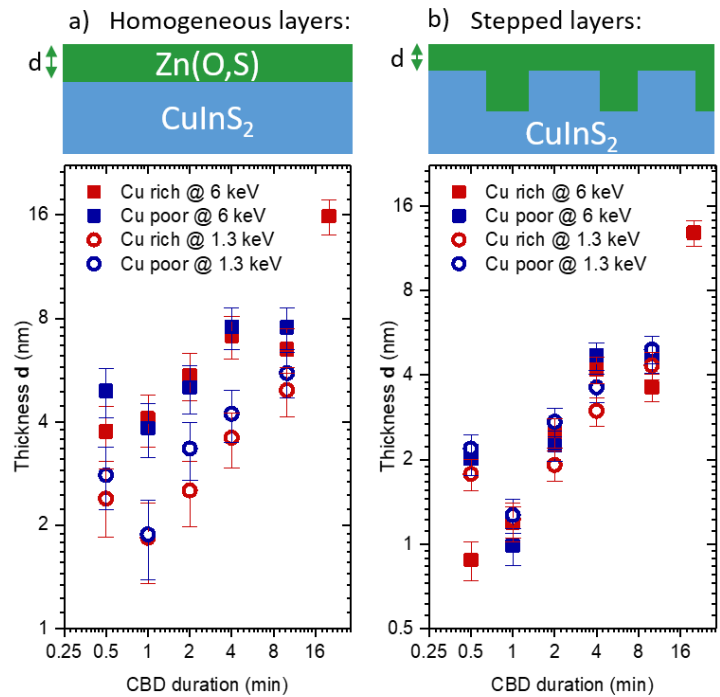

*Figure S12:* Calculated buffer thicknesses d of Zn(O,S) layers prepared using different chemical bath deposition (CBD) times on Cu rich and Cu poor CuInS$_2$ samples on a double log scale. Thicknesses were calculated assuming a homogeneous buffer layer (a) and a stepped buffer layer with 30% of the absorber surface represented by (buffer-filled) valleys (b), a sketch of the respective model is depicted on top of the displayed data. Thickness d was calculated from 1.3 keV and 6 keV measurements following formula 1 (a) and formula 2 (b).

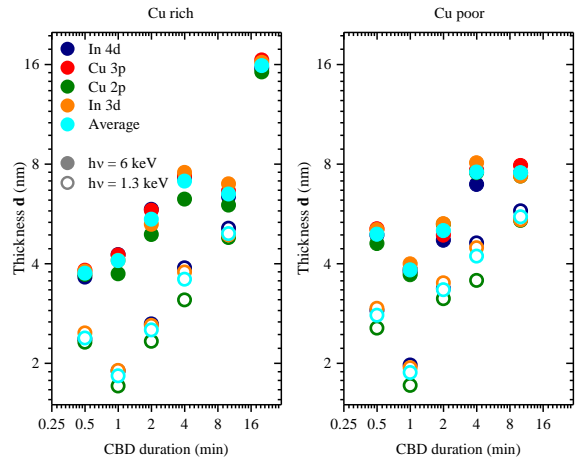

***Figure S13:*** *Calculated buffer thicknesses d of Zn(O,S) layers prepared using different chemical bath deposition (CBD) times on Cu rich (a) and Cu poor (b) CuInS$_2$ samples on a double log scale. Thicknesses d were calculated from 1.3 keV and 6 keV measurements assuming a homogeneous buffer layer following formula 1 in the manuscript. The thicknesses calculated from the In 4d, Cu 3p, Cu 2p, and In 3d are shown alongside an average value based on all used core levels.*

**HAXPES S 2p detail spectra:**

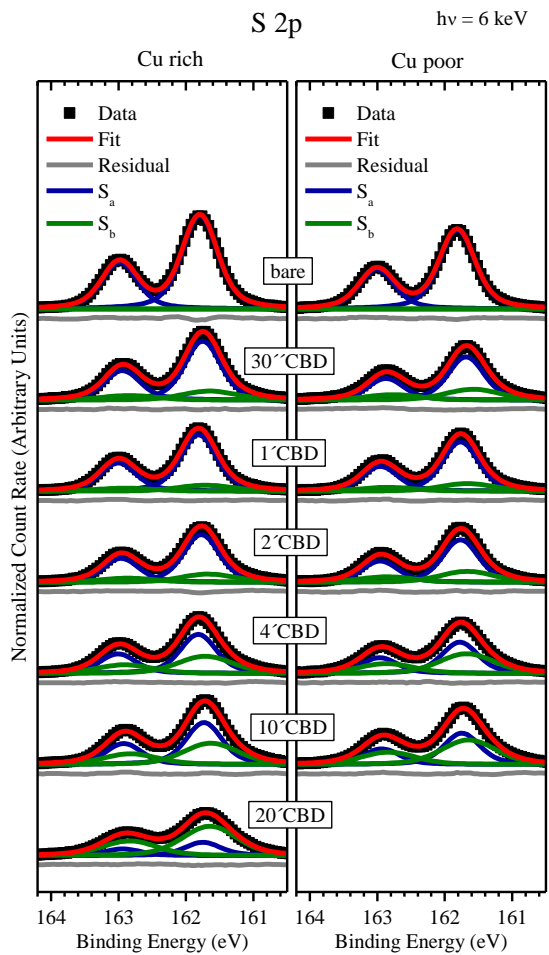

***Figure S14:*** *S 2p HAXPES (6 keV) spectra of Cu rich (left) and Cu poor (right) CuInS$_2$ absorbers with Zn(O,S) layers deposited using CBD for different durations (from 30 s to 20 min). Data are shown with a linear background subtracted. Respective fits using pairs of Voigt profiles to represent the respective doublets, are displayed along the data as well as the respective residuals. Spectra are vertically offset for clarity, as are residuals.*

Note, that the shape of S 2p component S$_a$ which was assigned to CuInS$_2$ and component S$_b$ which was assigned to ZnS were allowed to be different due to the expected different crystallinity of the materials. While the high-temperature processed CuInS$_2$ absorber is expected to be well ordered with a high degree of crystallinity, the low-temperature wet chemical deposited Zn(O,S) buffer is expected to have a low degree of crystallinity with even amorphous or nano-crystalline domains resulting in different bond angles and distances leading to a broadening of the Gaussian contribution to the Voigt profile. Also, component S$_a$ (attributed to the CuInS$_2$ absorber) was constrained to follow the same buffer growth – induced intensity attenuation as observed for the Cu and In related photoemission (see **Figure S13**).

**Zn(O,S) stoichiometry:**

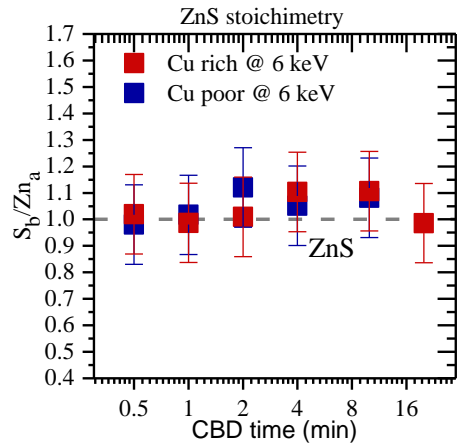

*Figure S15:* $S_b/Zn_a$ ratio as a measure of S:Zn ratio in the ZnS phase of the Zn(O,S) buffer. Ratio was calculated from spectra displayed in Figure 3 and Figure S14 considering corresponding photoionization cross sections, IMFPs and variations in electron analyzer transmission function.[50-52]

To corroborate the assumption of a stoichiometric ZnS and our Zn 3p fit model (Figure 3 in the manuscript), we used the information from the S 2p fit (**Figure S14**) to calculate $Zn_a/S_b$ ratios, corresponding to the Zn/S ratio of the ZnS phase of the Zn(O,S) buffer layer. It is close to one for all samples as shown in **Figure S15**, corroborating the used fit model and the assumption of a stoichiometric ZnS.

**CBM determination for the Zn(O,S) buffer on semi-log scale:**

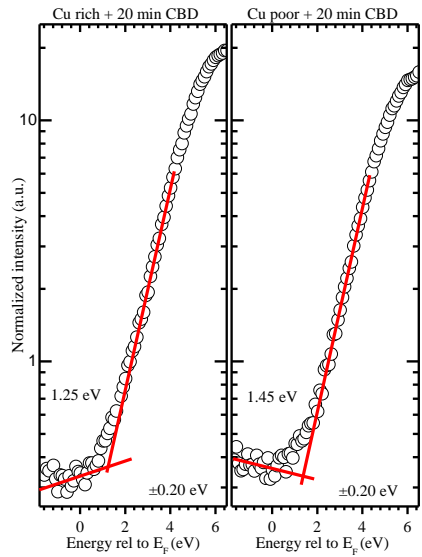

*Figure S16:* IPES spectra of the 20 min Zn(O,S) buffer layer deposited on Cu rich (left) and Cu poor (right) $CuInS_2$ absorbers. Data are shown on a semi-log scale together with an extrapolation of the main edge to derive CBM displayed as solid red line.

## Depth dependent band edges in absorber

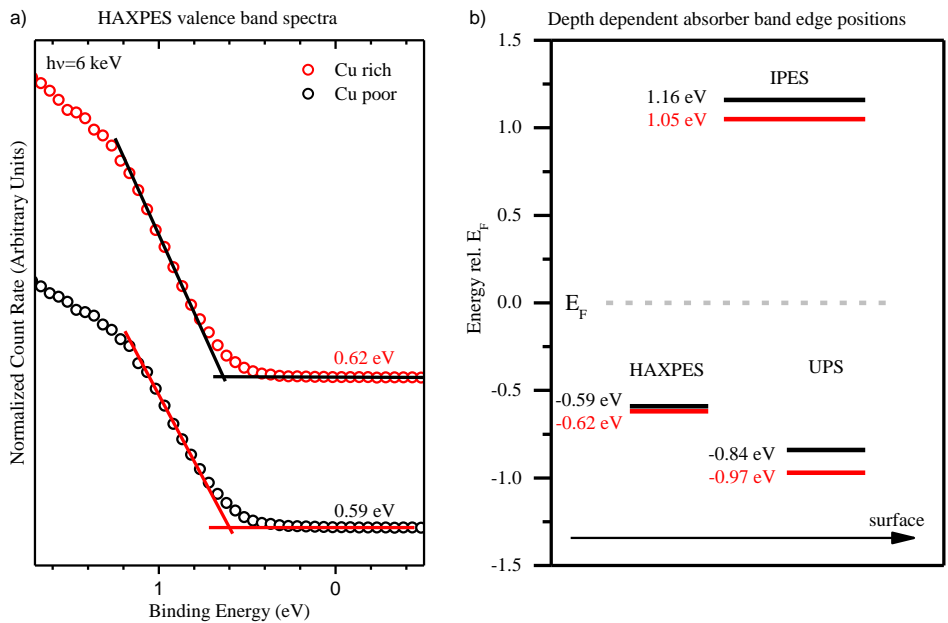

*Figure S17:* a): valence band spectra of the Cu rich and Cu poor bare absorbers determined using HAXPES (6 keV). Band edge and background are linearly extrapolated and the intersect is used to determine the band onset. Respective lines and onset values are given in the graph. b) band onsets of conduction and valence band derived by IPES, HAXPES, and UPS. As the less surface sensitive HAXPES measurements display a VBM closer to the Fermi level compared to the more sensitive UPS values, a VBM shift away from the Fermi level towards the surface can be concluded. Note, that even the HAXPES measurements are only probing the top 20-30 nm of the sample and the VBM might be even closer to the Fermi level further inside the bulk.

**Binding energy values for the considered core levels:**

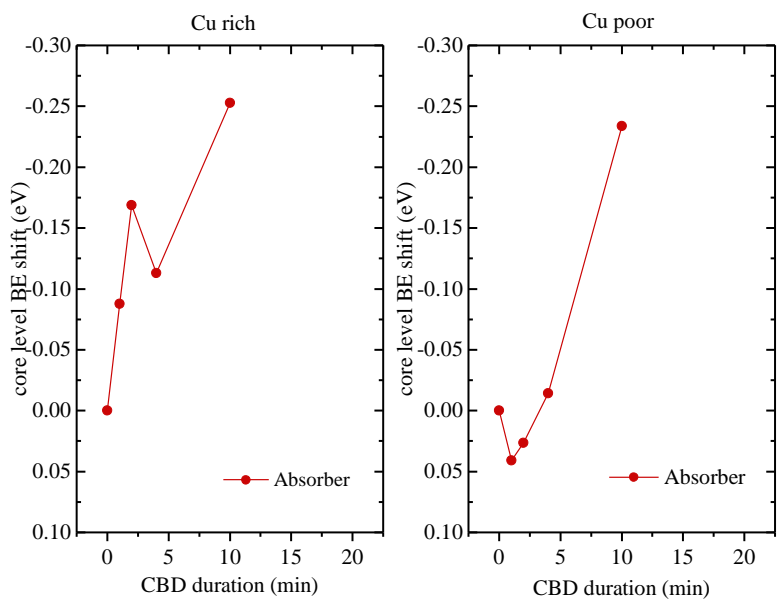

*Figure S18: Average core level shifts of the Cu 2p and In 3d peaks of the Cu-rich and Cu-poor CuInS$_2$ absorbers for different CBD times compared to the untreated absorbers.*

*Table S1:* Binding energies of the different contributions to the S 2p$_{3/2}$, Zn 3p$_{3/2}$ (both measured with 6 keV), Zn 2p$_{3/2}$, Cu 2p$_{3/2}$, and In 3d$_{3/2}$ (measured with 1.3 keV) core levels measured for a bare Cu rich CuInS$_2$ sample ('bare') and Zn(O,S) buffer layers CBD deposited within 30 s to 20 min, rounded to two digits.

| Cu rich | | Core level binding energy of respective sample (eV) | | | | | | |
|---|---|---|---|---|---|---|---|---|
| hν (keV) | Core level | Bare | 30´´ CBD | 1´CBD | 2´CBD | 4´CBD | 10´CBD | 20´CBD |
| 6 | S 2p$_a$ | 161.79 | 161.78 | 161.83 | 161.80 | 161.89 | 161.81 | 161.89 |
| 6 | S 2p$_b$ | | 161.51 | 161.56 | 161.53 | 161.62 | 161.54 | 161.62 |
| 6 | Zn 3p$_a$ | | 88.59 | 88.58 | 88.52 | 88.60 | 88.56 | 88.66 |
| 6 | Zn 3p$_b$ | | 89.22 | 89.21 | 89.15 | 89.23 | 89.19 | 39.29 |
| 1.3 | Zn 2p$_a$ | | 1022.02 | 1022.15 | 1021.95 | 1022.06 | 1022.07 | 1022.07 |
| 1.3 | Zn 2p$_b$ | | 1022.65 | 1022.78 | 1022.58 | 1022.69 | 1022.70 | 1022.70 |
| 1.3 | Cu 2p | 932.29 | 932.22 | 932.17 | 932.10 | 932.19 | 932.07 | |
| 1.3 | In 3d$_a$ | 452.34 | 452.25 | 452.28 | 452.19 | 452.21 | 452.06 | |
| 1.3 | In 3d$_b$ | 452.75 | 452.66 | 452.69 | 452.60 | 452.62 | 452.47 | |

*Table S2:* Binding energies of the different contributions in S 2p$_{3/2}$, Zn 3p$_{3/2}$ (both measured with 6 keV), Zn 2p$_{3/2}$, Cu 2p$_{3/2}$, and In 3d$_{3/2}$ (measured with 1.3 keV) core levels measured for a bare Cu rich CuInS$_2$ sample ('bare') and Zn(O,S) buffer layers CBD deposited within 30 s to 20 min, rounded to two digits.

| Cu poor | | Core level binding energy of respective sample (eV) | | | | | | |
|---|---|---|---|---|---|---|---|---|
| hν (keV) | Core level | Bare | 30´´ CBD | 1´CBD | 2´CBD | 4´CBD | 10´CBD | 20´CBD |
| 6 | S 2p$_a$ | 161.82 | 161.73 | 161.79 | 161.81 | 161.86 | 161.85 | 161.74 |
| 6 | S 2p$_b$ | | 161.46 | 161.52 | 161.54 | 161.59 | 161.58 | 161.47 |
| 6 | Zn 3p$_a$ | | 88.50 | 88.48 | 88.62 | 88.62 | 88.58 | 88.52 |
| 6 | Zn 3p$_b$ | | 89.13 | 89.11 | 89.25 | 89.25 | 89.21 | 89.15 |
| 1.3 | Zn 2p$_a$ | | 1021.94 | 1022.31 | 1022.28 | 1022.28 | 1022.08 | 1022.06 |

| 1.3 | Zn 2p$_b$ |        | 1022.57 | 1022.94 | 1022.91 | 1022.91 | 1022.71 | 1022.69 |
| --- | --------- | ------ | ------- | ------- | ------- | ------- | ------- | ------- |
| 1.3 | Cu 2p     | 932.29 | 932.08  | 932.32  | 932.30  | 932.29  | 932.08  |         |
| 1.3 | In 3d$_a$ | 452.37 | 452.11  | 452.41  | 452.41  | 452.34  | 452.10  |         |
| 1.3 | In 3d$_b$ | 452.78 | 452.52  | 452.82  | 452.82  | 452.75  | 452.51  |         |